\documentclass[11pt,a4paper]{article}
\usepackage[english]{babel}
\usepackage{amssymb,amsmath,amsfonts}
\usepackage[colorlinks=true,allcolors=blue]{hyperref}
\providecommand{\eqref}[1]{(\ref{#1})}

\newcommand{\MP}{M_\mathrm{P}}
\newcommand{\cA}{\mathcal{A}}
\newcommand{\cD}{\mathcal{D}}
\newcommand{\cL}{\mathcal{L}}
\newcommand{\cT}{\mathcal{T}}
\newcommand{\cV}{\mathcal{V}}
\newcommand{\dAlembert}{\square}
\DeclareMathOperator{\tr}{tr}
\DeclareMathOperator{\Tr}{Tr}
\newcommand{\nn}{\nonumber\\}
\newcommand{\ubar}[1]{\text{\b{$#1$}}}

\newcommand{\email}[1]{\footnote{E-mail: \href{mailto:#1}{#1}}}

\begin{document}
\title{Sakharov's induced gravity and the Poincar\'e gauge theory}

\author{Masud Chaichian,\email{masud.chaichian@helsinki.fi}
\ Markku Oksanen\email{markku.oksanen@helsinki.fi} \ and
Anca Tureanu\email{anca.tureanu@helsinki.fi}
\vspace{.8em}\\
\textit{Department of Physics, University of Helsinki, P.O. Box
64}\vspace{.2em}\\ \textit{FI-00014 Helsinki, Finland}
}
\date{}
\maketitle

\begin{abstract}
We explore Sakharov's seminal idea that gravitational dynamics is
induced by the quantum corrections from the matter sector. This was the
starting point of the view that gravity has an emergent origin, which
soon gained impetus due to the advent of black hole thermodynamics. In
the generalized framework of Riemann--Cartan spacetime with both
curvature and torsion, the induced gravitational action is obtained for
free nonminimally coupled scalar and Dirac fields. For a realistic
matter content, the induced Newton constant is obtained to be of the
magnitude of the ultraviolet cutoff, which implies that the cutoff is of
the order of the Planck mass.
Finally, we conjecture that the action for any gauge theory of gravity
at low energies can be induced by Sakharov's mechanism. This is
explicitly shown by obtaining the Poincar\'e gauge theory of gravity.
\end{abstract}

\vspace{1em}
\emph{Contribution to the Memorial Volume for Jacob Bekenstein}

\section{Introduction}
The standard approach for dealing with ultraviolet-divergent quantum
corrections in quantum field theory is renormalization. Its techniques
are well-developed mathematically, and the running of coupling constants
is verified experimentally for the Standard Model. In curved spacetime,
however, we encounter some problems. Particularly, the gravitational
couplings are radiatively unstable in the sense that the cosmological
constant and the Newton constant are extremely sensitive to any change
of the parameters of the matter sector or a change of the Wilsonian
cutoff scale of the matter action.

This problem could likely be solved by imposing constraints that keep
certain matter parameters and quantities of the effective action of
matter invariant under phase transitions and spontaneous symmetry
breaking. The implementation of such constraints requires new
physics beyond the Standard Model.
Here we explore an alternative and historically significant approach to
quantum corrections, which uses a cutoff scale that is
related to the observed gravitational couplings and offers an
illuminating view on the origin of gravity.

In Sakharov's approach to induced gravity \cite{Sakharov:1967pk},
classical gravity is considered to be induced by quantum corrections
from the matter sector.  At first matter fields are regarded to live on
a spacetime that is curved but with a nondetermined geometry.
The quantization of matter fields produces correction terms in the
action, which involve the curvature of spacetime. In particular,
one-loop corrections include the Einstein--Hilbert action of General
Relativity (GR). Finally, elevating the nondetermined metric of
spacetime to a dynamical variable turns the correction terms into a
gravitational action.
In other words, the regulated (but not renormalized) effective action of
matter on a curved spacetime is identified as the gravitational action
that determines the dynamics of spacetime geometry.
In order to avoid curvature terms of arbitrarily high orders to be
induced into the gravitational action, one usually assumes in
Sakharov's approach that quantum corrections beyond one-loop order are
somehow suppressed. Recall that already the squared Riemann
curvature terms, which are present in the one-loop effective action of
matter, generate new massive degrees of freedom that carry negative
energy \cite{Stelle:1976gc,Stelle:1977ry}, so-called ghosts, which
generate severe problems \cite{Kluson:2013hza}. When coupled to ordinary
fields, ghosts cause the system to evolve to an infinitely excited
state without a change in total energy. The inclusion of higher-order
curvature terms only makes the situation worse. Such terms are typically
present in the effective field theory of gravity, which is obtained as a
low-energy limit of various quantum theories.
In the induced gravitational action obtained in Sec.~\ref{sec:IG}, the
extra degrees of freedom are found to have masses around the Planck
mass, since the dimensionless coupling constants of the squared
curvature terms are smaller than one, and hence the characteristic
length scale in their Yukawa potentials
\cite{Stelle:1976gc,Stelle:1977ry} is the Planck length. Thus, the
effect of the extra degrees of freedom can be regarded to be negligible
at low energies and long distances. We assume that a consistent
description of quantum gravity and matter will eventually solve the
ghost problem. Therefore, in this work, we adopt the view that
higher-derivative contributions to the effective action can be ignored
in the low-energy regime.

The most important lesson of Sakharov's vision is that any fundamental
theory that includes or produces a curved spacetime manifold, on which a
quantum field theory of matter can be set up, necessarily produces
gravity as well. In this sense, gravity is an unavoidable and necessary
companion of quantum matter. The idea to induce gravity from quantum
effects and spontaneous symmetry breaking was further developed,
particularly by Adler and Zee \cite{Adler:1980bx,Adler:1980md,
Adler:1980pg,Adler:1982ri,Adler:1983,Zee:1978wi,Zee:1980sj,Zee:1983mj}.
A recent perspective on Sakharov's induced gravity is given in Ref.
\cite{Visser:2002ew}, which includes a discussion on different
interpretations of quantum corrections. The idea of induced gravity was
naturally expanded to the realm of quantum gravity, where it has been
used in particular to derive Einstein gravity as a low-energy
effective theory of scale-invariant (and asymptotically-free) quantum
field theory of gravity
\cite{Zee:1983mj,Smilga:1982se,Nepomechie:1983yq,Einhorn:2015lzy,
Einhorn:2016mws}.

When applied to matter fields, Sakharov's induced gravity is an early
representative of the emergent approach to gravity, where gravity or at
least gravitational dynamics is regarded to arise from a quantum theory
that does involve a gravitational interaction in its initial definition.
Sakharov's approach only addressess the induction of gravitational
dynamics by producing the gravitational action, and consequently the
field equations, but it does not produce spacetime, since the existence
of a curved spacetime manifold is presumed. Since then many people have
wondered whether spacetime too could be an emergent concept, and perhaps
even gravity as a whole might have an emergent origin. An intriguing
indication towards this view is the deep connection of gravity and
thermodynamics. It all began from Bekenstein's discovery of
the area law for black hole entropy \cite{Bekenstein:1972tm,
Bekenstein:1973ur, Bekenstein:1974ax,Bekenstein:1975tw}, which was
predated by observations that the horizon surface area and the
irreducible mass of black holes can never decrease in a classical
process \cite{Penrose:1971uk,Christodoulou:1970wf,
Christodoulou:1972kt,Hawking:1971tu}. The identification of horizon
surface area as entropy and surface gravity as temperature (both up to
a constant factor) quickly led to a full analogy between black hole
mechanics and thermodynamics \cite{Bardeen:1973gs}, and soon after to
Hawking's discovery of thermal radiation of black holes
\cite{Hawking:1974sw}. The universal upper limit on the entropy that can
be contained within a finite region of space which has a finite amount
of energy, namely, the Bekenstein bound
\cite{Bekenstein:1980jp,Bekenstein:1984vm,Schiffer:1989et,
Bekenstein:2004sh} means that a physical system with a finite energy in
a finite space is described by at most a certain finite amount of
information. A covariant generalization of the Bekenstein entropy bound
has been achieved \cite{Bousso:1999xy}, as well as a similar bound for
asymptotically de Sitter spacetimes \cite{Bousso:2000md}.
Black hole thermodynamics and the entropy bound were a major inspiration
for 't Hooft's proposal of the holographic principle
\cite{tHooft:1993dmi} and its subsequent string-theoretic interpretation
by Susskind \cite{Susskind:1994vu}. The gauge/gravity duality is the
most rigorous realization of the holographic principle
\cite{Maldacena:1997re}. It has been argued that since the black hole
entropy is, at least in part, an entanglement entropy
\cite{Bombelli:1986rw}, it would be most satisfactory if the
gravitational action is induced as Sakharov proposed, so that all black
hole entropy would be entanglement entropy
\cite{Jacobson:1994iw,Frolov:1997up}. Black hole thermodynamics (and the
holographic principle) has also had an influence on many other
approaches to understand the relation between gravity and
thermodynamics. The
Einstein equation has been derived locally on Rindler causal horizons as
a thermodynamic equation of state \cite{Jacobson:1995ab,Eling:2006aw}.
An extension of black hole thermodynamics to causal horizons has been
considered \cite{Jacobson:2003wv}. The holographic relation of bulk and
surface terms in gravitational actions \cite{Mukhopadhyay:2006vu} has
been used in arguing that the field equations of any diffeomorphism
invariant theory of gravity have a thermodynamic reinterpretation, and
showing that the equipartition of energy on the microscopic degrees of
freedom of a horizon can be used to derive gravity
\cite{Kothawala:2007em,Padmanabhan:2009kr,Padmanabhan:2010xh}. It has
even been proposed that gravity is an entropic force caused by changes
in the information associated with the positions of material bodies
\cite{Verlinde:2010hp}. The existence of gravitationally bound quantum
states\cite{Nesvizhevsky:2002ef} can be used to impose some conditions
on the fundamental microscopic theory behind entropic gravity
\cite{Chaichian:2011hc,Chaichian:2011xc}, which is presently unknown.
These results suggest a view of gravity and spacetime as emergent
concepts, which may have a thermodynamic origin.

The gauge theory approach to gravity has been highly influential ever
since the gauge invariance idea introduced by Weyl \cite{Weyl:1929fm}
for $U(1)$ was generalized to $SU(2)$ by Yang and Mills
\cite{Yang:1954ek} and to all semisimple Lie groups by Utiyama
\cite{Utiyama:1956sy}, who considered the gauging of the Lorentz group
for the first time. The first consistent gauge theory of gravity was
obtained by Kibble via gauging of the Poincar\'e group
\cite{Kibble:1961ba}, the symmetry group of the Minkowski spacetime,
which was used to derive the Einstein--Cartan--Sciama--Kibble theory of
gravity \cite{Sciama:1962,Kibble:1961ba}, but more generally yields a
Lagrangian that includes quadratic curvature and torsion terms
\cite{Obukhov:1987tz,Baekler:2010fr}.
From there on, gauge theories of every symmetry group related to gravity
have been proposed and explored, including the group of translations
\cite{Itin:2001bp}, the Weyl group (Poincar\'e group plus scale
transformations) \cite{Charap:1973fi}, the conformal and superconformal
groups \cite{Kaku:1977pa}, the affine group \cite{Lord:1978qz}, and so
on. See Ref.~\cite{Blagojevic:2012bc} for a review of various gauge
theories of gravity. The gauge theory approach has also been used in
attempts to understand the relation of gravity and quantum mechanics.
For example, several proposals for the gauge theory of gravity on
noncommutative spacetime have been considered, e.g.
\cite{Chamseddine:2000si,Chamseddine:2000zu,Cardella:2002pb,
Chaichian:2008pq}.

In this work, we consider Sakharov's induced gravity on a
Riemann--Cartan spacetime with both curvature and torsion. We shall
derive the induced gravitational action at one-loop order for free
scalar and Dirac fields. The mass scale that determines the induced
gravitational constants, especially the induced Newton constant
$G_\mathrm{ind}$, is the ultraviolet cutoff $\Lambda$ for the effective
action of matter fields. For gravity to have the observed strength,
$G_\mathrm{ind}^{-1}=8\pi\MP^2$, the ultraviolet cutoff $\Lambda$ has
to be comparable to the Planck mass $\MP$.
The effect of torsion is generally weak except when the density of
matter and spin is very high \cite{Hehl:1973,Hehl:1976kj}. When the
gravitational Lagrangian is the curvature scalar of the Riemann--Cartan
spacetime, $\cL=\frac{1}{2\kappa}\tilde R$, which gives the
Einstein--Cartan--Sciama--Kibble theory of gravity, torsion does not
propagate in vacuum. In more general theories, especially, in the
generic Poincar\'e gauge theory of gravity (PG) with a Lagrangian that
is quadratic in torsion and curvature, $\cL=\tilde R+T^2+\tilde R^2$,
torsion does
propagate in vacuum \cite{Adamowicz:1980qn,Chin:1983jt,Singh:1990yd}
(for recent development and further references, see
\cite{Blagojevic:2017wzf}), which may also improve the chances for its
detection in the future. For a discussion of several physical
implications of the torsion of spacetime, see
Refs.~\cite{Hehl:1973,Hehl:1976kj,Shapiro:2001rz}. The effect of the
induced gravitational terms with dimensionless couplings, including the
squared curvature terms, is found to be very small in the low-energy
regime, since their induced coupling constants depend on the logarithm
of the ultraviolet cutoff, which implies that those coupling constants
are of the order of one or lower.

Including torsion into the gravitational theory is highly appealing,
since then the intrinsic angular momentum of matter and gravity can be
naturally incorporated into the theory. In general, both torsion and
nonmetricity should be considered, in addition to curvature, in order to
obtain a comprehensive understanding of the nature of induced gravity in
the non-Riemannian setting. In this work, however, we confine the
treatment to spacetimes with a vanishing nonmetricity. This limitation
is motivated by our main goal, which is the emergence of PG as an
induced theory of gravity via Sakharov's mechanism. PG is a viable
alternative to GR, which has been studied extensively
\cite{Blagojevic:2012bc}. We show that the quantization of Dirac fields
on Riemann--Cartan spacetime induces the low-energy action of PG. The
high-energy part of the induced action is found to contain additional
terms compared to the action of PG. That is expectable, since the
effective action is not limited to contain only terms quadratic in the
field strengths, namely, in torsion and curvature. On a more general
spacetime with a non-metric compatible connection, a general
metric-affine gauge theory of gravity should be induced in the same way.

\section{Sakharov's approach}
\label{sec:Sakharov}
We generalize Sakharov's approach to Riemann--Cartan spacetime. As in
Riemannian spacetime, the procedure can be considered to consist of
five steps:
\begin{enumerate}
\item Assume a Lorentzian spacetime manifold. The geometric notions of
Riemann--Cartan spacetime can be derived by applying Einstein's
Equivalence Principle to a Dirac spinor \cite{Blagojevic:2012bc}, which
can be considered to describe a neutron in a gravitational field,
instead of applying it to a point mass as in GR.
\item Leave the dynamics of the geometry undetermined, i.e., consider
it as an arbitrary classical background and do not define an action for
gravity.
\item Quantize matter fields and determine their effective action. In
the language of Feynman diagrams, the one-loop effective action
represents the sum of all one-loop diagrams coupled to an arbitrary
number of external gravitons. In this visualization, gravitons refer to
small perturbations of the geometric background fields around a fixed
background. Regularize the effective action, and obtain the
contribution of the background geometry to the action.
\item Elevate the geometric background fields to gravitational
variables. We consider two possible choices for the variables:
the vierbein and the spin connection (a first-order formalism), or the
metric and the torsion (a second-order formalism).
\item Identify the regulated one-loop effective action as the leading
contribution to gravity. The gravitational action induced at one-loop
order consists of contributions to vacuum energy, curvature terms up to
second order and torsion terms up to fourth order.
\end{enumerate}

Sakharov's approach, and many subsequent induced gravity approaches,
involve several problems:
\begin{enumerate}
\item The induced Newton constant is not guaranteed to be unique and
positive. In general, the Feynman amplitudes for stress-energy tensor
operators are complex, and should be continued analytically. This
problem also appears in the approaches\cite{Adler:1982ri,Zee:1978wi}
where gravity is induced via symmetry breaking \cite{David:1983tx}. Some
ways to addresss the problem has been proposed \cite{Novozhilov:1991nm}.
A further ambiguity is caused by the choice of a regularization method,
since different methods imply different quantum corrections.

An ultraviolet-finite induced Newton constant can be obtained by
including both scalar fields and spin-1/2 fields and by imposing
fine-tuning constraints \cite{Frolov:1997up} on the numbers, masses and
couplings of the scalar and spin-1/2 fields.
% Then the induced Newton constant is independent of the regularization
% method.
Unfortunately, the masses of the constituent fields have to be
comparable with the Planck mass, which causes problems in the presence
of gravity, since quantum gravity effects should become significant at
high energies.
\item The introduction of dimensional parameters that determine the
scale of gravitational couplings is not fully convincing. In our
simplistic approach, cutoff regularization is used to set the mass
parameter of the induced theory of gravity to be of the order of the
Planck mass. Alternatively, one could use any other regularization
method, for example, Pauli--Villars regulators with sufficiently high
masses. If the scale of gravitational couplings is set by the masses of
fundamental fields, one generally requires fields whose masses are
comparable with the Planck, which is problematic.
\item The elevation of the geometric notions to gravitational variables
after the quantization of matter fields has no physical motivation
(other than that it works). This problem is a consequence of the fact
that Sakharov's approach does not address the emergence of spacetime
geometry but rather only the emergence of gravitational dynamics.
\end{enumerate}

\section{Geometric definitions}%: torsion and curvature}
\label{sec:TR}
The covariant derivatives associated with the connection involving
torsion $\tilde{\Gamma}$ and the torsionless connection $\Gamma$ are
denoted by $\tilde\nabla$ and $\nabla$, respectively.
The connections on Riemann--Cartan spacetime are defined to be metric
compatible. Namely, the nonmetricity tensor is assumed to vanish
throughout this work, $Q_{\mu\nu\rho}=\tilde\nabla_\mu g_{\nu\rho}=0$.
Relaxing the metric-compatibility would lead to more general spacetime
geometries, e.g., Weyl--Cartan \cite{Charap:1973fi} or metric-affine
\cite{Lord:1978qz,Hehl:1994ue}, which are not considered here. Greek
indices ($\mu,\nu,\ldots$) refer to a coordinate basis, and Latin
indices ($a,b,\ldots$) refer to an orthonormal noncoordinate basis.

The connection coefficient
$\tilde{\Gamma}_{\mu\nu}^{\phantom{\mu\nu}\rho}$ can be written as the
sum of the Christoffel symbol $\Gamma_{\mu\nu}^{\phantom{\mu\nu}\rho}$
and the contortion tensor $K_{\mu\nu}^{\phantom{\mu\nu}\rho}$,
\begin{equation}\label{tildeGamma}
 \tilde{\Gamma}_{\mu\nu}^{\phantom{\mu\nu}\rho}
 =\Gamma_{\mu\nu}^{\phantom{\mu\nu}\rho}
 +K_{\mu\nu}^{\phantom{\mu\nu}\rho},
\end{equation}
where the contortion tensor $K_{\mu\nu}^{\phantom{\mu\nu}\rho}$ is
defined by the torsion $T_{\mu\nu}^{\phantom{\mu\nu}\rho}$,
\begin{equation}
T_{\mu\nu}^{\phantom{\mu\nu}\rho}=
2\tilde{\Gamma}_{[\mu\nu]}^{\phantom{[\mu\nu]}\rho}
=\tilde{\Gamma}_{\mu\nu}^{\phantom{\mu\nu}\rho}
-\tilde{\Gamma}_{\nu\mu}^{\phantom{\nu\mu}\rho},
\end{equation}
as
\begin{equation}\label{contortion}
 K_{\mu\nu}^{\phantom{\mu\nu}\rho}=\frac{1}{2}\left(
 T_{\mu\nu}^{\phantom{\mu\nu}\rho}
 +T_{\phantom{\rho}\mu\nu}^\rho
 +T_{\phantom{\rho}\nu\mu}^\rho \right).
\end{equation}
In the second-order formalism of gravity, the independent gravitational
variables can be chosen as the metric and the torsion, which determine
the connection \eqref{tildeGamma}.
Torsion can be decomposed into three irreducible components: the trace
vector $\cV_\mu=T_{\mu\nu}^{\phantom{\mu\nu}\nu}$, the axial vector
$\cA^\mu=T_{\nu\rho\sigma}\epsilon^{\nu\rho\sigma\mu}$, and the tensor
component $\cT_{\mu\nu}^{\phantom{\mu\nu}\rho}$ with vanishing vector
and axial vector parts, $\cT_{\mu\nu}^{\phantom{\mu\nu}\nu}=0$ and
$\cT_{\nu\rho\sigma}\epsilon^{\nu\rho\sigma\mu}=0$. Then, in
four-dimensional spacetime, we have
\begin{equation}
 T_{\mu\nu\rho}=\frac{1}{3}\left( \cV_\mu g_{\nu\rho}
 -\cV_\nu g_{\mu\rho} \right) -\frac{1}{6}\epsilon_{\mu\nu\rho\sigma}
 \cA^\sigma +\cT_{\mu\nu\rho}.
\end{equation}
The components of torsion couple to matter fields in different ways,
which is discussed for scalar and Dirac fields in Sec.~\ref{sec:IG}.
The contortion tensor is expressed in terms of the components of
torsion as
\begin{equation}\label{K.decom}
 K_{\mu\nu}^{\phantom{\mu\nu}\rho}=\frac{1}{3}\left(
 g_{\mu\nu}\cV^\rho -\delta_\mu^{\ \rho}\cV_\nu \right)
 -\frac{1}{12}\epsilon_{\mu\nu\lambda\sigma}g^{\lambda\rho}\cA^\sigma
 +\frac{1}{2}\left( \cT_{\mu\nu}^{\phantom{\mu\nu}\rho}
 +\cT_{\phantom{\rho}\mu\nu}^\rho
 +\cT_{\phantom{\rho}\nu\mu}^\rho \right),
\end{equation}
% \begin{equation}\label{K.decom}
%  K_{\mu\nu\rho}=\frac{1}{3}\left( g_{\mu\nu}\cV_\rho
%  -g_{\mu\rho}\cV_\nu \right)
%  -\frac{1}{12}\epsilon_{\mu\nu\rho\sigma}\cA^\sigma
%  +\frac{1}{2}\left( \cT_{\mu\nu\rho} -\cT_{\nu\rho\mu}
%  +\cT_{\rho\mu\nu} \right)
% \end{equation}
and we note that $K_{\mu\nu}^{\phantom{\mu\nu}\mu}=
-K_{\mu\phantom{\mu}\nu}^{\phantom{\mu}\mu}=-\cV_\nu$ and
$K_{\mu\nu\rho}\epsilon^{\mu\nu\rho\sigma}=\frac{1}{2}\cA^\sigma$.

The curvature tensor of the connection \eqref{tildeGamma} can be
written in terms of the curvature tensor of the torsionless connection
and the contortion tensor as\footnote{Tensor indexes in brackets are
antisymmetrized, $a_{[\mu}b_{\nu]}=\frac{1}{2}(a_\mu b_\nu-a_\nu
b_\mu)$, and indexes between vertical lines are excluded from the
antisymmetrization, $a_{[\mu|\rho}b_{|\nu]}=\frac{1}{2}(a_{\mu\rho}b_\nu
-a_{\nu\rho}b_\mu)$.}
\begin{equation}\label{Riem.rel}
 \tilde{R}_{\mu\nu\phantom\rho\sigma}^{\phantom{\mu\nu}\rho}=
 R_{\mu\nu\phantom\rho\sigma}^{\phantom{\mu\nu}\rho}
 +2\nabla_{[\mu}^{} K_{\nu]\sigma}^{\phantom{\nu\sigma}\rho}
 +2K_{[\mu|\lambda}^{\phantom{[\mu|\lambda}\rho}
 K_{|\nu]\sigma}^{\phantom{|\nu]\sigma}\lambda},
\end{equation}
where
\begin{equation}\label{tildeR}
 \tilde{R}_{\mu\nu\phantom\rho\sigma}^{\phantom{\mu\nu}\rho}=
 2\partial_{[\mu}^{}
 \tilde{\Gamma}_{\nu]\sigma}^{\phantom{\nu]\sigma}\rho}
 +2\tilde{\Gamma}_{[\mu|\lambda}^{\phantom{[\mu|\lambda}\rho}
 \tilde{\Gamma}_{|\nu]\sigma}^{\phantom{|\nu]\sigma}\lambda},
\end{equation}
and $R_{\mu\nu\phantom\rho\sigma}^{\phantom{\mu\nu}\rho}$ is defined
similarly in terms of $\Gamma_{\mu\nu}^{\phantom{\mu\nu}\rho}$.
We may also express the curvature tensor of the torsionless connection
as
\begin{equation}\label{Riem.rel2}
 R_{\mu\nu\phantom\rho\sigma}^{\phantom{\mu\nu}\rho}=
 \tilde{R}_{\mu\nu\phantom\rho\sigma}^{\phantom{\mu\nu}\rho}
 -2\tilde{\nabla}_{[\mu}^{} K_{\nu]\sigma}^{\phantom{\nu\sigma}\rho}
 -T_{\mu\nu}^{\phantom{\mu\nu}\lambda}
 K_{\lambda\sigma}^{\phantom{\lambda\sigma}\rho}
 +2K_{[\mu|\lambda}^{\phantom{[\mu|\lambda}\rho}
 K_{|\nu]\sigma}^{\phantom{|\nu]\sigma}\lambda},
\end{equation}
where everything on the right-hand side is defined in terms of the
connection with torsion.
Note that the order of tensor indices on curvature and torsion is such
that the commutator of two covariant derivatives is written as
\begin{equation}
 [\tilde\nabla_\mu,\tilde\nabla_\nu]V^\rho=
 \tilde{R}_{\mu\nu\phantom\rho\sigma}^{\phantom{\mu\nu}\rho}V^\sigma
 -T_{\mu\nu}^{\phantom{\mu\nu}\sigma}\tilde\nabla_\sigma V^\rho,
\end{equation}
and the order will persist when the vierbein formalism and the spin
representation are considered.
The relations between the Ricci tensors, $\tilde{R}_{\mu\nu}=
\tilde{R}_{\mu\rho\phantom\rho\nu}^{\phantom{\mu\rho}\rho}$ and
$R_{\mu\nu}=R_{\mu\rho\phantom\rho\nu}^{\phantom{\mu\rho}\rho}$, and
the scalar curvatures, $\tilde{R}=g^{\mu\nu}\tilde{R}_{\mu\nu}$ and
$R=g^{\mu\nu}R_{\mu\nu}$, can be obtained from the relations
\eqref{Riem.rel} and \eqref{Riem.rel2}. For the scalar curvatures we get
\begin{equation}
\tilde{R}=R-2\nabla_\mu\cV^\mu+\frac{2}{3}\cV_\mu\cV^\mu
-\frac{1}{24}\cA_\mu\cA^\mu
-\frac{1}{2}\cT_{\mu\nu\rho}\cT^{\mu\nu\rho},
\end{equation}
or the other way around,
\begin{equation}
R=\tilde{R}+2\tilde\nabla_\mu\cV^\mu+\frac{4}{3}\cV_\mu\cV^\mu
+\frac{1}{24}\cA_\mu\cA^\mu
+\frac{1}{2}\cT_{\mu\nu\rho}\cT^{\mu\nu\rho}.
\end{equation}

In the first-order formalism, which is particularly used in the gauge
theory approach to gravity, the independent variables are the vierbein
$e^a_{\phantom{a}\mu}$ and the spin connection
$\tilde{\omega}_{\mu\phantom{a}b}^{\phantom{\mu}a}$, or the
corresponding one-forms, the coframe
$\theta^a=e^a_{\phantom{a}\mu}dx^\mu$ and the connection
$\tilde{\omega}^a_{\phantom{a}b}=
\tilde{\omega}_{\mu\phantom{a}b}^{\phantom{\mu}a}dx^\mu$.
These variables are the gauge fields that are required to ensure
gauge invariance of the action.
Additionally, in a metric-affine gauge theory like PG, we assume the
presence of a metric, $g_{ab}\theta^a\otimes\theta^b$, which is here
taken to be an orthonormal coframe, $g_{ab}=\eta_{ab}$ (the Minkowski
metric), so that the metric in a coordinate basis is defined as
\begin{equation}
g_{\mu\nu}=\eta_{ab}e^a_{\phantom{a}\mu}e^b_{\phantom{b}\nu}.
\end{equation}
We consider PG, which is based on the Lorentz connection,
$\tilde{\omega}^{ab}=\tilde{\omega}^{[ab]}$, i.e., on an antisymmetric
linear connection. More general gauge theories of gravity require a
more general linear connection.\footnote{Extending the local Poincar\'e
group with local scale transformations would require the connection to
have a nonvanishing trace component, which leads to a gauge theory of
gravity with Weyl--Cartan geometry \cite{Charap:1973fi}. A general
metric-affine gauge theory of gravity is based on the local affine
group, and it requires an unrestricted linear connection
\cite{Lord:1978qz,Hehl:1994ue}.}
In the presence of a metric, it is possible to decompose the spin
connection into a torsionless connection and a contortion component
\begin{equation}\label{omegarelation}
 \tilde{\omega}_\mu^{\phantom{\mu}ab}=\omega_\mu^{\phantom{\mu}ab}
 +K_{\mu}^{\phantom{\mu}\nu\rho}e^{a}_{\phantom{a}\nu}
 e^{b}_{\phantom{b}\rho},
\end{equation}
which consists of the torsionless part $\omega_\mu^{\phantom{\mu}ab}$
and the part proportional to contortion. The spin connection is related
to the connection in a coordinate frame by the so-called tetrad
postulate
\begin{equation}
\tilde\nabla_\mu e^{a}_{\phantom{a}\nu}=\partial_\mu
e^{a}_{\phantom{a}\nu}+\tilde\omega_{\mu\phantom{a}b}^{\phantom{\mu}a}
e^{b}_{\phantom{b}\nu} -\tilde\Gamma_{\mu\nu}^{\phantom{\mu\nu}\rho}
e^{a}_{\phantom{a}\rho}=0,
\end{equation}
which can be equivalently written as
\begin{equation}
\tilde\omega_{\mu\phantom{a}b}^{\phantom{\mu}a}
=\left( \tilde\Gamma_{\mu\nu}^{\phantom{\mu\nu}\rho}
e^{a}_{\phantom{a}\rho} -\partial_\mu e^{a}_{\phantom{a}\nu} \right)
e_{b}^{\phantom{b}\nu}.
\end{equation}

The torsion two-form is defined in terms of $e^a_{\phantom{a}\mu}$ and
$\tilde{\omega}_{\mu\phantom{a}b}^{\phantom{\mu}a}$ as the exterior
covariant derivative of the orthonormal coframe,
\begin{equation}\label{torsion.PG}
T^a=d\theta^a+\tilde{\omega}^a_{\phantom{a}b}\wedge\theta^b
=\frac{1}{2}T_{\mu\nu}^{\phantom{\mu\nu}a}
dx^\mu\wedge dx^\nu,
\end{equation}
where the components are defined as
\begin{equation}
T_{\mu\nu}^{\phantom{\mu\nu}a}=2\partial_{[\mu}^{}
e^a_{\phantom{a}\nu]}
+2\tilde{\omega}_{[\mu|\phantom{a}b}^{\phantom{[\mu|}a}
e^b_{\phantom{b}|\nu]}.
\end{equation}
The curvature two-form is defined as
\begin{equation}\label{tildeR.PG}
\tilde{R}^a_{\phantom{a}b}=d\tilde{\omega}^a_{\phantom{a}b}
+\tilde{\omega}^a_{\phantom{a}c}\wedge\tilde{\omega}^c_{\phantom{c}b}
=\frac{1}{2}\tilde{R}_{\mu\nu\phantom{a}b}^{\phantom{\mu\nu}a}
dx^\mu\wedge dx^\nu,
\end{equation}
where the components are written as
\begin{equation}\label{tildeRmunuab}
\tilde{R}_{\mu\nu\phantom{a}b}^{\phantom{\mu\nu}a}
=2\partial_{[\mu}^{}\tilde{\omega}_{\nu]\phantom{a}b}^{\phantom{\nu]}a}
+2\tilde{\omega}_{[\mu|\phantom{a}c}^{\phantom{[\mu|}a}
\tilde{\omega}_{|\nu]\phantom{c}b}^{\phantom{|\nu]}c}.
\end{equation}
The curvature $R_{\mu\nu\phantom{a}b}^{\phantom{\mu\nu}a}$ for the
torsionless connection $\omega_{\mu\phantom{a}b}^{\phantom{\mu}b}$ is
defined similarly as \eqref{tildeRmunuab}.
In a coordinate basis, the components of the curvature are given as
\begin{equation}
\tilde{R}_{\mu\nu\phantom\rho\sigma}^{\phantom{\mu\nu}\rho}
=\tilde{R}_{\mu\nu\phantom{a}b}^{\phantom{\mu\nu}a}
e_a^{\phantom{a}\rho}e^b_{\phantom{b}\sigma}.
\end{equation}
When expressed entirely in the orthonormal frame $\{e_a\}$
(and the coframe $\{\theta^a\}$), where
$e_a=e_a^{\phantom{a}\mu}\partial_\mu$, the components of the torsion
and the curvature are given as
\begin{align}
\tilde{R}_{ab\phantom{c}d}^{\phantom{ab}c}&=2e_{[a}^{}
\tilde\omega_{b]\phantom{c}d}^{\phantom{b]}c}
+2\tilde\omega_{[a|\phantom{c}e}^{\phantom{[a|}c}
\tilde\omega_{|b]\phantom{e}d}^{\phantom{|b]}e}
-c_{ab}^{\phantom{ab}e}\tilde\omega_{e\phantom{c}d}^{\phantom{e}c},\\
T_{ab}^{\phantom{ab}c}&=2\tilde\omega_{[a\phantom{c}b]}^{\phantom{[a}c}
-c_{ab}^{\phantom{ab}c},
\end{align}
where $\tilde\omega_{a\phantom{c}b}^{\phantom{a}c}
=e_a^{\phantom{a}\mu}\tilde\omega_{\mu\phantom{c}b}^{\phantom{\mu}c}$,
and they involve the anholonomity of the basis:
\begin{equation}
[e_a,e_b]=c_{ab}^{\phantom{ab}c}e_c,\quad
c_{ab}^{\phantom{ab}c}=
2e_{[a|}^{\phantom{[a|}\mu}\partial_\mu^{}e_{|b]}^{\phantom{|b]}\nu}
e^c_{\phantom{c}\nu}.
% =\left( e_a^{\phantom{a}\mu}\partial_\mu e_b^{\phantom{b}\nu}
% -e_b^{\phantom{b}\mu}\partial_\mu e_a^{\phantom{a}\nu}\right)
% e^c_{\phantom{c}\nu}.
\end{equation}

\section{Induced gravitational action from quantized matter fields}
\label{sec:IG}

One can considers matter fields that are coupled to curvature (and
in the present case also to torsion) minimally, although there is no
physical principle that would require such a restriction. In some
cases, e.g. for scalar fields, nonminimal couplings are necessary in
order to achieve renormalizability. Here our goal is not
renormalization, but rather derivation of the one-loop effective action
that is regulated but not renormalized. That way the effective action
can be regarded as the origin of gravity. Therefore, the presence of
nonminimal couplings is not necessary for the present approach, but we
shall consider them for the sake of generality.

\subsection{Scalar fields}\label{sec:scalar}
First consider a free real scalar field $\phi$ on four-dimensional
Riemann--Cartan spacetime. We can obtain the minimally-coupled
Lagrangian from a free-field Lagrangian on Minkowski spacetime with the
replacement $\partial_\mu\rightarrow\tilde\nabla_\mu$. The usual
free-field Lagrangian on Minkowski spacetime with the kinetic part
$\frac{1}{2}\eta^{\mu\nu}\partial_\mu\phi\partial_\nu\phi$ gives the
same Lagrangian as in Riemannian spacetime, since
$\tilde\nabla_\mu\phi=\nabla_\mu\phi=\partial_\mu\phi$, and hence the
torsion does not appear in it.
% \begin{equation}
%  \frac{1}{2}g^{\mu\nu}\tilde\nabla_\mu\phi\tilde\nabla_\nu\phi
%  =\frac{1}{2}g^{\mu\nu}\nabla_\mu\phi\nabla_\nu\phi.
% \end{equation}
Expressing the kinetic part of the Lagrangian in Minkowski spacetime as
$-\frac{1}{2}\phi\eta^{\mu\nu}\partial_\mu\partial_\nu\phi$ leads to a
different Lagrangian with a coupling to the vector component of
torsion as
\begin{equation}
 -\frac{1}{2}\phi\tilde\dAlembert\phi=-\frac{1}{2}\phi\left(
 \dAlembert\phi +\cV^\mu\nabla_\mu\phi \right),
\end{equation}
where
\begin{equation}
 \tilde\dAlembert=g^{\mu\nu}\tilde\nabla_\mu\tilde\nabla_\nu,\quad
 \dAlembert=g^{\mu\nu}\nabla_\mu\nabla_\nu.
\end{equation}
Thus, the minimal replacement rule does not lead to a unique scalar
field Lagrangian in Riemann--Cartan spacetime. Unlike in a Riemannian
spacetime, the form of the initial Lagrangian in flat spacetime matters.
Since it would be a limited viewpoint to consider only the coupling
to the vector component of torsion, we will include other torsion terms
as well. Hence, we consider the parity-conserving free field Lagrangian
with all nonminimal coupling terms
\begin{equation}\label{L_scalar}
 \cL=\frac{1}{2}\left( g^{\mu\nu}\nabla_\mu\phi\nabla_\nu\phi
 -m^2\phi^2 -\sum_{i=1}^5 \xi_i P_i\phi^2 \right),
\end{equation}
where $\xi_i$ are dimensionless coupling constants and the corresponding
even-parity curvature and torsion terms are
\begin{equation}\label{P_i}
 P_1=R,\quad P_2=\nabla_\mu \cV^\mu,\quad P_3=\cV_\mu \cV^\mu,\quad
 P_4=\cA_\mu\cA^\mu,\quad P_5=\cT_{\mu\nu\rho}\cT^{\mu\nu\rho}.
\end{equation}
The Lagrangian \eqref{L_scalar} could alternatively be written in terms
of $\tilde\nabla$, $\tilde{R}$ and the components of torsion.

Note that we choose to write the Lagrangian in terms of the torsionless
covariant derivative and its curvature, since it makes the calculations
more convenient by enabling the use of techniques developed for quantum
fields on Riemannian spacetime (see the monographs
\cite{Birrell:1982ix,Parker:2009uva}). That is, the metric and the
torsion are regarded as the background fields. Note, however, that the
gravitational variables are not chosen yet. They will be chosen later,
after the quantum effective action has been obtained. Similar to the
case of spacetime without torsion, where only the first nonminimal
coupling $\xi R\phi^2$ appears, the nonminimal couplings would be
necessary for renormalization \cite{Buchbinder:1985ux}.  Naturally,
including pseudoscalar fields and/or complex scalar fields would allow
further nonminimal coupling terms, which would again be necessary for
renormalization \cite{Buchbinder:1985ux}. Recall, however, that our
present goal is not renormalization.

The action is defined by the Lagrangian \eqref{L_scalar} as
\begin{equation}
 S=\int d^4x\sqrt{-g}\,\cL=-\frac{1}{2}\int d^4x\sqrt{-g}\,\phi D\phi,
\end{equation}
where
\begin{equation}\label{D}
 D=\dAlembert+m^2+\sum_{i=1}^5 \xi_i P_i.
\end{equation}
The one-loop effective action is defined
\cite{Avramidi:2000bm,Parker:2009uva} in terms of $D$ as
\begin{equation}
 S_\mathrm{eff}^{(1)}=\frac{i}{2}\ln\det\left( l^2D \right)
 =\frac{i}{2}\Tr\ln\left( l^2D \right),
\end{equation}
where the parameter $l$ with dimension of length was introduced for
dimensional reasons, so that $l^2D$ is dimensionless. Since we consider
only one-loop corrections, and hence there is no need to keep track of
higher powers of $\hbar$, we have set $\hbar=1$, along with $c=1$. We
can use the identity
\begin{equation}\label{lnD/D0}
\ln\left( \frac{D}{D_0} \right)=-\int_0^\infty \frac{d\tau}{\tau}
 \left[ \exp(-i\tau D)-\exp(-i\tau D_0) \right],
\end{equation}
where $D_0$ is the operator for a suitable reference background
$(g_0,\tilde\Gamma_0)$,\footnote{Both operators $D$
and $D_0$ are considered to have a small negative imaginary part in
order to avoid divergence of the integration in \eqref{lnD/D0}.}
in order to obtain
\begin{equation}\label{S_eff}
 S_\mathrm{eff}^{(1)}=S_\mathrm{eff,0}^{(1)}
 -\frac{i}{2}\Tr\int_0^\infty \frac{d\tau}{\tau}
 \left[ \exp(-i\tau D)-\exp(-i\tau D_0) \right],
\end{equation}
where $S_\mathrm{eff,0}^{(1)}$ is the one-loop effective action for the
reference background. On a noncompact spacetime the reference
background has to be chosen so that the physical action for induced
gravity $S_\mathrm{phys}^{(1)}=S_\mathrm{eff}^{(1)}
-S_\mathrm{eff,0}^{(1)}$ is well defined. We shall consider a compact
spacetime for simplicity. This does not limit the generality of the
derivation. The contribution of a reference background can be easily
included into the induced action afterwards, in case one needs the
result for a noncompact spacetime. Thence, instead of associating the
operator $D_0$ with a reference background, we can freely choose it to
be proportional to an identity operator as $D_0=l^{-2}I$, so that we
obtain
\begin{equation}\label{S_eff.I}
S_\mathrm{eff}^{(1)}=
-\frac{i}{2}\Tr\int_0^\infty \frac{d\tau}{\tau}\exp(-i\tau D)
+\frac{i}{2}\lim_{\epsilon\rightarrow0+}\int_0^\infty
\frac{ds}{s} \exp(-is(1-i\epsilon)) \Tr I.
\end{equation}
In units of mass, the dimensions of the above constant $l$ and the
integration variables $\tau$ and $s=l^{-2}\tau$ are $[l]=-1$,
$[\tau]=-2$ and $[s]=0$. The second term in \eqref{S_eff.I} is a
constant, which is irrelevant dynamically, since it does not depend on
$D$ or on any geometric quantities. Therefore, in the following, we drop
the constant term and write
\begin{equation}\label{effact0.def}
 S_\mathrm{eff}^{(1)}= -\frac{i}{2}\Tr\int_0^\infty \frac{d\tau}{\tau}
 \exp(-i\tau D).
\end{equation}

We express the operator in the effective action in terms of a kernel
function $K(\tau;x,y)$, which is defined as
\begin{equation}
 \exp(-i\tau D)\phi(x)=\int d^4y\sqrt{-g} K(\tau;x,y)\phi(y).
\end{equation}
The kernel satisfies a Schr\"odinger-like equation
\begin{equation}
 i\frac{d}{d\tau}K(\tau;x,y)=DK(\tau;x,y),
\end{equation}
with the boundary condition
\begin{equation}
 K(\tau=0;x,y)=I\delta(x,y),
\end{equation}
where $I$ is the identity operator/matrix for the fields or field
components. We are working with a spacetime of Lorentzian signature so
that the kernel corresponds to a heat kernel with imaginary time
\cite{Avramidi:2000bm,Vassilevich:2003xt}. Strictly speaking, the heat
kernel technique is mathematically well defined only in Euclidean
signature, when the squared distance $(x-y)^2$ of points is positive
definite. The kernel discussed here should be regarded as an analytic
continuation of the heat kernel, and it is only useful for finding the
local contributions (in the limit $y\rightarrow x$) to the effective
action. The trace of the operator in \eqref{effact0.def} is given by
\begin{equation}\label{TrexpD}
 \Tr\exp(-i\tau D)=\int d^4x\sqrt{-g}\tr K(\tau;x,x),
\end{equation}
where in the right-hand side the trace is taken over the field degrees
of freedom.\footnote{When a field with several components or several
fields are involved, the operator $D$ and its kernel $K$ are
matrix-valued.}
The (divergent) gravitational terms that we are interested in come from
the zero end of the integral over $\tau$.
% Hence, the upper limit of the integral can be taken to be finite
% $\tau_0$.
We use the local series expansion of the kernel
\cite{Birrell:1982ix,Avramidi:2000bm,Parker:2009uva, Vassilevich:2003xt}
\begin{equation}\label{Kernel}
K(\tau;x,x)=\frac{i}{(4\pi i\tau)^2}\sum_{n=0}^\infty (i\tau)^n A_n(x),
\end{equation}
where $A_n(x)$ are constructed from the geometric quantities and
parameters involved in $D$, i.e., from the covariant derivative, the
curvature, the torsion, and the masses and couplings of the fields.
We assume that the spacetime has no boundary. If the spacetime had a
boundary, we would have to include boundary terms into the series
expansion of the operator \eqref{TrexpD}. In the presence of boundaries,
each term in the series expansion except the zeroth term involves an
additional boundary contribution \cite{Vassilevich:2003xt}, and the
series expansion \eqref{Kernel} also involves half-integer terms,
$n=\frac{1}{2},\frac{3}{2},\ldots$, which are purely boundary terms.

Since the operator \eqref{D} is readily in the Laplace form
\begin{equation}\label{Doperator}
 D=\dAlembert+B,
\end{equation}
where the (endomorphism) term $B$ does not involve
derivatives,\footnote{We mean that $B$ does not involve derivative
operators acting on the field. Of course, $B$ itself involves
derivatives of the gravitational fields, namely, in the present case the
curvature and torsion terms in \eqref{B0}.}
one obtains the first three terms of the kernel \eqref{Kernel}
as\footnote{See Refs.\cite{Avramidi:2000bm,Gilkey:1975iq} and references
therein for different techniques of derivation for the kernel
coefficients, and e.g. Refs.\cite{Birrell:1982ix,Parker:2009uva,
Vassilevich:2003xt} for its use.}
\begin{align}
 A_0&=I,\label{A_0}\\
 A_1&=\frac{1}{6}RI-B,\label{A_1}\\
 A_2&=\left( \frac{1}{180}R^{\mu\nu\rho\sigma}R_{\mu\nu\rho\sigma}
 -\frac{1}{180}R^{\mu\nu}R_{\mu\nu} +\frac{1}{72}R^2
 -\frac{1}{30}\dAlembert R \right) I \nn
 &\quad +\frac{1}{2}B^2 -\frac{1}{6}RB +\frac{1}{6}\dAlembert B
 +\frac{1}{12}W^{\mu\nu}W_{\mu\nu},\label{A_2}
\end{align}
where
\begin{equation}\label{W_munu}
 W_{\mu\nu}=[\nabla_\mu,\nabla_\nu].
\end{equation}
For a single real scalar field, the identity is of course
one-dimensional, $I=1$, \eqref{W_munu} vanishes, $W_{\mu\nu}=0$, and
\begin{equation}\label{B0}
 B=m^2+\sum_{i=1}^5 \xi_i P_i.
\end{equation}

Which terms of the expansion of the kernel \eqref{Kernel} appear in the
regularized one-loop effective action depends on the chosen
regularization method. For example, dimensional regularization involves
only the term $A_2$, which contains quadratic curvature and torsion
terms. In that sense, it is a too powerful regularization method for our
purposes. We choose to use the cutoff regularization, since it involves
all the given terms \eqref{A_0}--\eqref{A_2}. The lower limit of the
integral over $\tau$ is cut off at $\Lambda^{-2}$ for the three
divergent terms, where the ultraviolet cutoff parameter $\Lambda$ has
the dimension of mass, $[\Lambda]=1$. For the third term with $A_2$ the
upper limit of the integral is cut off at $\tau_0=\epsilon^{-2}$,
where $\epsilon$ is an infrared cutoff. We obtain the kernel expansion
of the cutoff-regularized one-loop effective action as
\begin{equation}\label{effact0}
\begin{split}
S_\mathrm{eff}^{(1)}&=\frac{1}{32\pi^2}\sum_{n=0}^\infty
\int_{0\;\mathrm{or}\;\Lambda^{-2}}^{\infty\;\mathrm{or}\;\epsilon^{-2}}
d(i\tau) (i\tau)^{n-3} \int d^4x\sqrt{-g} \tr A_n(x) \\
% &=-\frac{1}{32\pi^2}\bigg\vert_{\Lambda^{-2}}^{\Lambda^{-2}_0}
% \frac{\tau^{-2}}{2} \int d^4x\sqrt{-g}A_0
% -\frac{1}{32\pi^2}\bigg\vert_{\Lambda^{-2}}^{\Lambda^{-2}_0}\tau^{-1}
% \int d^4x\sqrt{-g}A_1\\
% &\quad+\frac{1}{32\pi^2}\bigg\vert_{\Lambda^{-2}}^{\Lambda^{-2}_0}
% \ln\tau \int d^4x\sqrt{-g}A_2 +\text{ultraviolet-finite terms}\\
&=\frac{\Lambda^4}{64\pi^2}\int d^4x\sqrt{-g}\tr A_0
+\frac{\Lambda^2}{32\pi^2}\int d^4x\sqrt{-g}\tr A_1 \\
&\quad+\frac{\ln(\Lambda/\epsilon)}{16\pi^2}\int d^4x\sqrt{-g}\tr A_2
+\text{ultraviolet-finite terms}.
\end{split}
\end{equation}
The term of order $\Lambda^4$, i.e., the quartic divergence, is given as
$\tr A_0=1$, which contributes only to the vacuum energy.
The term of order $\Lambda^2$ in the one-loop effective action
\eqref{effact0} is given as
\begin{equation}\label{TrA_1}
 \tr A_1=-m^2 +\left( \frac{1}{6}-\xi_1 \right)R
 -\sum_{i=2}^5\xi_iP_i.
\end{equation}
The term proportional to the logarithm of the cutoff contains the
second-order gravitational terms: quadratic curvature terms and torsion
terms up to fourth power. The term $\frac{1}{2}B^2$ of the contribution
\eqref{A_2} proportional to $\ln(\Lambda/\epsilon)$ also contains the
first-order terms $m^2\sum_{i=1}^5\xi_iP_i$. In total, we have
\begin{equation}\label{trA_2.scalar}
\begin{split}
 \tr A_2&=\frac{1}{180}R^{\mu\nu\rho\sigma}R_{\mu\nu\rho\sigma}
 -\frac{1}{180}R^{\mu\nu}R_{\mu\nu} +\frac{1}{72}R^2
 -\frac{1}{30}\dAlembert R \\
 &\quad +\frac{1}{2}\left( m^2+\sum_{i=1}^5 \xi_i P_i \right)^2
 -\frac{1}{6}R\left( m^2+\sum_{i=1}^5 \xi_i P_i \right)
 +\frac{1}{6}\sum_{i=1}^5 \xi_i \dAlembert P_i \\
 &=\frac{1}{180}R^{\mu\nu\rho\sigma}R_{\mu\nu\rho\sigma}
 -\frac{1}{180}R^{\mu\nu}R_{\mu\nu}
 +\frac{1}{2}\left( \frac{1}{36}-\frac{\xi_1}{3}+\xi_1^2 \right)R^2 \\
 &\quad +\frac{1}{6}\left( \xi_1-\frac{1}{5} \right)\dAlembert R
 +\left( \xi_1-\frac{1}{6} \right) R\sum_{i=2}^5 \xi_i P_i
 +\frac{1}{2}\left( \sum_{i=2}^5 \xi_i P_i \right)^2\\
 &\quad +\frac{1}{6}\sum_{i=2}^5 \xi_i \dAlembert P_i
 +\left( \xi_1-\frac{1}{6} \right)m^2R
 +m^2\sum_{i=2}^5 \xi_i P_i +\frac{1}{2}m^4,
\end{split}
\end{equation}
where the sums in the latter expression are taken over the torsion
terms, $P_i$ with $i=2,\ldots,5$.

The Gauss-Bonnet-Chern term
\begin{equation}\label{GBC}
G=R_{\alpha\beta\gamma\delta}R_{\mu\nu\rho\sigma}
\epsilon^{\alpha\beta\mu\nu}\epsilon^{\gamma\delta\rho\sigma}
=R^{\mu\nu\rho\sigma}R_{\mu\nu\rho\sigma}
- 4R^{\mu\nu}R_{\mu\nu} + R^2,
\end{equation}
whose integral is a topological invariant in four-dimensional
spacetime, can be used to absorb the Riemann tensor squared term
from the effective action.

Alternatively, we could write the effective action in terms of the
connection $\tilde\nabla$ with torsion and its curvature \eqref{tildeR}
by using the relations between the two connections \eqref{tildeGamma}
and their curvature tensors \eqref{Riem.rel2}. That would be the
appropriate way, if one chooses the first-order formalism, where
$e^a_{\phantom{a}\mu}$ and
$\tilde{\omega}_{\mu\phantom{a}b}^{\phantom{\mu}a}$ are the independent
variables of the induced gravitational action. Such a case is
considered in Sec.~\ref{sec:inducedPG}, where the induced action for PG
is obtained.

\subsection{Spin-1/2 fields}\label{sec:spin1/2}
Next we consider a Dirac spinor field $\psi$. The Dirac field can
couple to the vector and the axial vector components of torsion.
Naturally, these couplings are of the same form as the couplings for any
other vector and axial vector fields. In particular, the vector
component $\cV_\mu$ couples to the Dirac spinor in the same way as the
electromagnetic field.
In the Dirac action, spin-1/2 fields do not couple to curvature or
to higher-order torsion invariants \eqref{P_i} due to dimensional
reasons, since the mass dimensions are $[\psi]=\frac{3}{2}$,
$[Riemann]=2$ and $[Torsion]=1$.\footnote{We do
not consider coupling constants with negative mass dimensions or
couplings to fractional or negative powers of curvature and
torsion.}

The Hermitian Lagrangian for a free minimally coupled Dirac field is
written as
\begin{equation}\label{L_1/2,min}
 \cL_{\text{Dirac}}^{\text{min.}}=\frac{i}{2} \left(
 \bar{\psi}\ubar{\gamma}^\mu\tilde\nabla_\mu\psi
 -\tilde\nabla_\mu\bar{\psi}\ubar{\gamma}^\mu\psi \right)
 -m\bar{\psi}\psi,
\end{equation}
where the spacetime-dependent $\gamma$ matrices are defined as
$\ubar{\gamma}^\mu=e_{a}^{\phantom{a}\mu}\gamma^a$, where
$e_{a}^{\phantom{a}\mu}$ is the inverse of the vierbein
$e^{a}_{\phantom{a}\mu}$.
In four-dimensional spacetime, the constant $\gamma$ matrices satisfy
\begin{equation}
 \{\gamma^a,\gamma^b\}=2\eta^{ab}I,\quad
  (\gamma^0)^2=I,\quad  (\gamma^i)^2=-I,\quad
 (\gamma^a)^\dagger=\gamma^0\gamma^a\gamma^0,
\end{equation}
where $a=0,1,2,3$, $i=1,2,3$, $(\eta^{ab})=\mathrm{diag}(1,-1,-1,-1)$
and $I$ is the  four-dimensional identity matrix. The fifth $\gamma$
matrix is defined as
\begin{equation}
 \gamma^5=i\gamma^0\gamma^1\gamma^2\gamma^3,
\end{equation}
and it satisfies
\begin{equation}
 (\gamma^5)^2=I,\quad (\gamma^5)^\dagger=\gamma^5,\quad
 \{\gamma^5,\gamma^a\}=0.
\end{equation}
The covariant derivative for the spinor and its conjugate
$\bar\psi=\psi^\dagger\gamma^0$ is defined by
\begin{equation}\label{tildenabla.spinor}
\begin{split}
 \tilde\nabla_\mu\psi&=\partial_\mu\psi
 +\frac{i}{2}\tilde{\omega}_\mu^{\phantom{\mu}ab}\Sigma_{ab}\psi,\\
 \tilde\nabla_\mu\bar\psi&=\partial_\mu\bar\psi
 -\frac{i}{2}\bar\psi\tilde{\omega}_\mu^{\phantom{\mu}ab}\Sigma_{ab},
\end{split}
\end{equation}
where $\Sigma_{ab}=-\frac{i}{4}[\gamma_a,\gamma_b]$ satisfies the
Lie algebra of $SO(1,3)$.
Likewise, the covariant derivative without torsion is
\begin{equation}\label{nabla.spinor}
\begin{split}
 \nabla_\mu\psi&=\partial_\mu\psi
 +\frac{i}{2}\omega_\mu^{\phantom{\mu}ab}\Sigma_{ab}\psi,\\
 \nabla_\mu\bar\psi&=\partial_\mu\bar\psi
 -\frac{i}{2}\bar\psi\omega_\mu^{\phantom{\mu}ab}\Sigma_{ab}.
\end{split}
\end{equation}
The action for the minimally coupled Lagrangian \eqref{L_1/2,min} can
be written in two forms, either in terms of $\tilde\nabla$ or $\nabla$,
as
\begin{equation}\label{S_1/2,min}
\begin{split}
 S_{\text{Dirac}}^{\text{min.}}&=\int d^4x\sqrt{-g}\bar{\psi}\left(
 i\ubar{\gamma}^\mu\tilde\nabla_\mu
 -\frac{i}{2}\ubar{\gamma}^\mu\cV_\mu -m \right)\psi \\
 &=\int d^4x\sqrt{-g}\bar{\psi}\left(
 i\ubar{\gamma}^\mu\nabla_\mu +\frac{1}{8}\ubar{\gamma}^\mu
 \gamma^5\cA_\mu -m \right)\psi,
\end{split}
\end{equation}
where the boundary surface term coming from an integration by parts is
assumed to vanish, $\int d^4x\partial_\mu(\sqrt{-g}\bar{\psi}
\ubar{\gamma}^\mu\psi)=0$. According to the second expression of
\eqref{S_1/2,min} a minimally coupled spinor couples only to the
axial vector component of torsion. We shall consider a more general
Dirac field that couples to both the vector and axial components of
torsion.

Written in terms of the torsionless covariant derivative, a
nonminimally coupled Dirac field has the action
% \begin{equation}
% S_{1/2}=\int d^4x\sqrt{-g}\left(
% i\bar{\psi}\ubar{\gamma}^\mu\nabla_\mu\psi
% -\alpha\bar{\psi}\ubar{\gamma}^\mu\cV_\mu\psi
% -\beta\bar{\psi}\ubar{\gamma}^\mu\cA_\mu\gamma^5\psi
% -m\bar{\psi}\psi \right),
% \end{equation}
\begin{equation}\label{S_1/2}
\begin{split}
S_{\text{Dirac}}^{\text{non-min.}}&=\int d^4x\sqrt{-g}\bar{\psi}
D_{1/2}\psi,\\
D_{1/2}&=i\ubar{\gamma}^\mu\cD_\mu -m,\\
\cD_\mu&=\nabla_\mu +i\alpha_1\cV_\mu
+i\alpha_2\gamma^5\cA_\mu,
\end{split}
\end{equation}
where $\alpha_1$ and $\alpha_2$ are dimensionless coupling constants
for the vector and the axial vector components of torsion,
respectively. The minimally coupled case \eqref{L_1/2,min} corresponds
to the couplings $\alpha_1=0$, $\alpha_2=-\frac{1}{8}$.

The one-loop effective action is defined as
\begin{equation}\label{effact1/2.def}
 S_\mathrm{eff}^{(1)}=-i\ln\det(lD_{1/2}),
\end{equation}
where the minus sign and the factor of two compared to the spin-0 case
\eqref{effact0.def} come from the anticommuting and complex-valued
nature of the Dirac field, respectively. Since the Dirac operator
$D_{1/2}$ is a first-order differential operator, we cannot use the
technique of Sec.~\ref{sec:scalar} directly. However, we can square the
operator $D_{1/2}$ as follows (for a more detailed proof, see
\cite{DeBerredoPeixoto:2001qm}). First we define a modification of the
operator $D_{1/2}$ as
\begin{equation}
D_{1/2}^{*}=-i\ubar{\gamma}^\mu\cD_\mu -m.
\end{equation}
In even-dimensional spacetime, the matrix $\gamma^5$, which is
Hermitian and satisfies $(\gamma^5)^2=I$, anticommutes with
$\ubar{\gamma}^\mu$. Hence,
$\gamma^5D_{1/2}^{*}\gamma^5=D_{1/2}$,\footnote{In even-dimensional
spacetime, the matrices $\{\ubar{\gamma}^\mu\}$ and
$\{-\ubar{\gamma}^\mu\}$ form equivalent representations of the Clifford
algebra (see e.g. \cite{Parker:2009uva}).}
and we obtain
\begin{equation}
\det(lD_{1/2}^{*})=\det(l\gamma^5D_{1/2}^{*}\gamma^5)
=\det(lD_{1/2}).
\end{equation}
Therefore, the operator $D_{1/2}$ in the effective action
\eqref{effact1/2.def} can be squared as
\begin{equation}
 S_\mathrm{eff}^{(1)}=-\frac{i}{2}\ln\left[\det(lD_{1/2})\right]^2
 =-\frac{i}{2}\ln\det(l^2D_{1/2}D_{1/2}^{*}).
\end{equation}
% assuming that there is no net chiral anomaly.
Hence, the operator $D$ in the effective action of a Dirac spinor is
written as
\begin{equation}\label{Dspinor}
 D\equiv D_{1/2}D_{1/2}^{*}=(\ubar{\gamma}^\mu\cD_\mu)^2 +m^2.
\end{equation}
The operator $D$ in \eqref{Dspinor} is not of the Laplace type for any
of the derivatives $\cD$, $\nabla$ or $\tilde\nabla$.
We can see this by expanding the first term in \eqref{Dspinor} as
\begin{equation}
\begin{split}
 (\ubar{\gamma}^\mu\cD_\mu)^2&=g^{\mu\nu}\cD_\mu\cD_\nu
 -2i\alpha_2\ubar{\gamma}^\mu\ubar{\gamma}^\nu\gamma^5\cA_\mu\cD_\nu
 +\frac{1}{2}\ubar{\gamma}^\mu\ubar{\gamma}^\nu[\cD_\mu,\cD_\nu],
\end{split}
\end{equation}
where the last term can be written as
\begin{equation}
 \frac{1}{2}\ubar{\gamma}^\mu\ubar{\gamma}^\nu[\cD_\mu,\cD_\nu]
 =\frac{1}{4}R +\frac{i}{4}\alpha_1[\ubar{\gamma}^\mu,
 \ubar{\gamma}^\nu] V_{\mu\nu}
 +\frac{i}{4}\alpha_2[\ubar{\gamma}^\mu,\ubar{\gamma}^\nu]\gamma^5
 A_{\mu\nu},
\end{equation}
which involves the following second-rank tensors\footnote{In a more
general setting, the vector $\cV_\mu$ and the axial vector $\cA_\mu$
could contain not only the vector components of torsion but also gauge
fields in some representation of a given gauge group. In that case,
$V_{\mu\nu}$ and $A_{\mu\nu}$ would be extended to include the field
strength tensors of the gauge fields.}
\begin{equation}
\begin{split}
 V_{\mu\nu}&=\nabla_\mu\cV_\nu-\nabla_\nu\cV_\mu,\\
 A_{\mu\nu}&=\nabla_\mu\cA_\nu-\nabla_\nu\cA_\mu.
\end{split}
\end{equation}
Therefore, the operator \eqref{Dspinor} contains a first-order
derivative term. However, for any partial differential operator $D$ that
contains first-order and second-order partial derivatives, with the
second-order term taking the form $g^{\mu\nu}\partial_\mu\partial_\nu$,
there exists \cite{Gilkey:1975iq} a unique connection $\hat\nabla$ and a
unique endomorphism $B$ for which the operator takes the Laplace form
\begin{equation}\label{Laplacian.spinor}
 D=\hat{\dAlembert} +B,\quad
 \hat{\dAlembert}=g^{\mu\nu}\hat\nabla_\mu\hat\nabla_\nu.
\end{equation}
That connection consists of the Riemannian connection and a gauge
(bundle) part. In the present case \eqref{Dspinor}, the operator
\eqref{Laplacian.spinor} is given by
\begin{equation}
\begin{split}
 \hat\nabla_\mu&=\cD_\mu -i\alpha_2\ubar{\gamma}^\nu
 \ubar{\gamma}_\mu\gamma^5\cA_\nu \\
 &=\nabla_\mu +i\alpha_1\cV_\mu
 +\frac{i}{2}\alpha_2\bigl[\ubar{\gamma}_\mu,\ubar{\gamma}_\nu\bigr]
 \gamma^5\cA^\nu
\end{split}
\end{equation}
% or
% \begin{equation}
%  w_\mu=\frac{1}{8}\omega_\mu^{\phantom{\mu}ab}[\gamma_a,\gamma_b]
%  +i\alpha_1\cV_\mu
%  +\frac{i}{2}\alpha_2\bigl[\ubar{\gamma}_\mu,\ubar{\gamma}_\nu\bigr]
%  \gamma^5\cA^\nu,
% \end{equation}
% so that
% \begin{multline}
%  \hat{\dAlembert}=g^{\mu\nu}\cD_\mu\cD_\nu
%  -2i\alpha_2\ubar{\gamma}^\mu\ubar{\gamma}^\nu\gamma^5\cA_\mu\cD_\nu
%  -i\alpha_2\left( \gamma^5\nabla_\mu\cA^\mu
% -\frac{1}{4}[\ubar{\gamma}^\mu,
%  \ubar{\gamma}^\nu]]\gamma^5A_{\mu\nu} \right) \\
% +2\alpha_2^2\cA_\mu\cA^\mu,
% \end{multline}
and
\begin{equation}\label{B1/2}
 B=\left( \frac{1}{4}R -2\alpha_2^2\cA_\mu\cA^\mu +m^2 \right)I
 +\frac{i}{4}\alpha_1[\ubar{\gamma}^\mu, \ubar{\gamma}^\nu] V_{\mu\nu}
 +i\alpha_2\gamma^5\nabla_\mu\cA^\mu.
\end{equation}
The tensor $W_{\mu\nu}$ for the connection $\hat\nabla$
is given by
\begin{equation}\label{W1/2}
\begin{split}
W_{\mu\nu}
% &=[\hat\nabla_\mu,\hat\nabla_\nu]\\
% &=-\frac{1}{8} R_{\mu\nu}^{\phantom{\mu\nu}ab}[\gamma_a,\gamma_b]
% +i\alpha_1 V_{\mu\nu} +\frac{i}{2}\alpha_2\bigl(
% \bigl[\ubar{\gamma}_\nu,\ubar{\gamma}_\rho\bigr]\gamma^5
% \nabla_\mu\cA^\rho \\
% &\quad -\bigl[\ubar{\gamma}_\mu,\ubar{\gamma}_\rho\bigr]\gamma^5
% \nabla_\nu\cA^\rho \bigr) -\frac{1}{4}\alpha_2^2\bigl(
% \bigl[\ubar{\gamma}_\mu,\ubar{\gamma}_\rho\bigr]\cA^\rho
% \bigl[\ubar{\gamma}_\nu,\ubar{\gamma}_\sigma\bigr]\cA^\sigma \\
% &\quad -\bigl[\ubar{\gamma}_\nu,\ubar{\gamma}_\rho\bigr]\cA^\rho
% \bigl[\ubar{\gamma}_\mu,\ubar{\gamma}_\sigma\bigr]\cA^\sigma
% \bigr)\\
&=-\frac{1}{8} R_{\mu\nu\rho\sigma} [\ubar{\gamma}^\rho,
\ubar{\gamma}^\sigma]
+i\alpha_1 V_{\mu\nu} +i\alpha_2\gamma^5 A_{\mu\nu}\\
&\quad +i\alpha_2\gamma^5\ubar{\gamma}_\rho \left(
\ubar{\gamma}_\mu\nabla_\nu\cA^\rho
-\ubar{\gamma}_\nu\nabla_\mu\cA^\rho \right)\\
&\quad-\alpha_2^2\left( \ubar{\gamma}_\mu\ubar{\gamma}_\rho
\ubar{\gamma}_\nu\ubar{\gamma}_\sigma
-\ubar{\gamma}_\nu\ubar{\gamma}_\rho
\ubar{\gamma}_\mu\ubar{\gamma}_\sigma
\right) \cA^\rho\cA^\sigma.
\end{split}
\end{equation}

The kernel expansion and regularization of the effective action are
performed in a way identical to that of the scalar field case. Only now
the operator $D$, the kernel $K$, the identity $I$ and the tensor
$W_{\mu\nu}$ are four-dimensional square matrices. The tensor
$W_{\mu\nu}$ is given for the spinor in \eqref{W1/2}. The
cutoff-regularized effective action has the same expression as for a
scalar field \eqref{effact0} but with an opposite sign,
\begin{equation}\label{effact1/2}
\begin{split}
 S_\mathrm{eff}^{(1)}
 &=-\frac{\Lambda^4}{64\pi^2}\int d^4x\sqrt{-g}\tr A_0
 -\frac{\Lambda^2}{32\pi^2}\int d^4x\sqrt{-g}\tr A_1 \\
 &\quad-\frac{\ln(\Lambda/\epsilon)}{16\pi^2}\int d^4x\sqrt{-g}\tr A_2
 +\text{ultraviolet-finite terms}.
\end{split}
\end{equation}
Then we evaluate the traces of the terms \eqref{A_0}--\eqref{A_2}
of the kernel expansion with \eqref{B1/2} and \eqref{W1/2}. The first
two terms of the effective action \eqref{effact1/2} are given by
\begin{align}
 \tr A_0&=4,\\
 \tr A_1&=-\frac{1}{3}R+8\alpha_2^2\cA_\mu\cA^\mu -4m^2,
\end{align}
using $\tr I=4$ and $\tr B=R -8\alpha_2^2\cA_\mu\cA^\mu +4m^2$.
% Calculation of the trace of $A_2$ is more laborious. First we obtain
% \begin{equation}
%  \tr B^2=4\left( \frac{1}{4}R -2\alpha_2^2\cA_\mu\cA^\mu +m^2
% \right)^2
%  +2\alpha_1^2 V^{\mu\nu}V_{\mu\nu} -4\alpha_2^2(\nabla_\mu\cA^\mu)^2,
% \end{equation}
% where we also used \eqref{trgamma.2}, \eqref{trgamma.4} and
% \eqref{trgamma.2.5} in the forms
% $\tr([\ubar{\gamma}^\mu,\ubar{\gamma}^\nu])=0$,
% $\tr([\ubar{\gamma}^\mu,\ubar{\gamma}^\nu]
% [\ubar{\gamma}^\rho,\ubar{\gamma}^\sigma])
% =-16(g^{\mu\rho}g^{\nu\sigma}
% -g^{\mu\sigma}g^{\nu\rho})$ and
% $\tr([\ubar{\gamma}^\mu,\ubar{\gamma}^\nu]\gamma^5)=0$. Then
% \begin{equation}
% \begin{split}
%  \tr(RB)&=R^2 -8\alpha_2^2R\cA_\mu\cA^\mu +4m^2R,\\
%  \tr\dAlembert B&=\dAlembert R
% -8\alpha_2^2\,\dAlembert(\cA_\mu\cA^\mu).
% \end{split}
% \end{equation}
% Lastly, we obtain the trace of the square of \eqref{W1/2} as
% \begin{equation}
% \begin{split}
%  \tr(W^{\mu\nu}W_{\mu\nu})&=-\frac{1}{2}R^{\mu\nu\rho\sigma}
%  R_{\mu\nu\rho\sigma} +8\alpha_2^2 R \cA_\mu\cA^\mu
%  -4\alpha_1^2 V^{\mu\nu}V_{\mu\nu}\\
%  &\quad +8\alpha_2^2 A^{\mu\nu}A_{\mu\nu}
%  -96\alpha_2^4(\cA_\mu\cA^\mu)^2
%  +24\alpha_2^2(\nabla_\mu\cA^\mu)^2 \\
%  &\quad  +16\alpha_2^2 \nabla_\mu\left( \cA^\nu\nabla_\nu\cA^\mu
%  -\cA^\mu\nabla_\nu\cA^\nu \right),
% \end{split}
% \end{equation}
% where most of the identities from Appendix~\ref{app:gamma} were
% used.
Finally, after some lengthy algebra, we obtain the term proportional to
$\ln(\Lambda/\epsilon)$ in the effective action \eqref{effact1/2} as
\begin{equation}\label{trA_2.spinor}
\begin{split}
 \tr A_2
%  &=\frac{4}{180}R^{\mu\nu\rho\sigma}R_{\mu\nu\rho\sigma}
%  -\frac{4}{180}R^{\mu\nu}R_{\mu\nu} +\frac{4}{72}R^2
%  -\frac{4}{30}\dAlembert R \\
%  &\quad +2\left( \frac{1}{4}R -2\alpha_2^2\cA_\mu\cA^\mu +m^2\right)^2
%  +\alpha_1^2 V^{\mu\nu}V_{\mu\nu} -2\alpha_2^2(\nabla_\mu\cA^\mu)^2 \\
%  &\quad -\frac{1}{6}R^2 +\frac{4}{3}\alpha_2^2R\cA_\mu\cA^\mu
%  -\frac{2}{3}m^2R \\
%  &\quad +\frac{1}{6}\dAlembert R
%  -\frac{4}{3}\alpha_2^2\,\dAlembert(\cA_\mu\cA^\mu) \\
%  &\quad -\frac{1}{24}R^{\mu\nu\rho\sigma}R_{\mu\nu\rho\sigma}
%  +\frac{2}{3}\alpha_2^2 R \cA_\mu\cA^\mu
%  -\frac{1}{3}\alpha_1^2 V^{\mu\nu}V_{\mu\nu}\\
%  &\quad +\frac{2}{3}\alpha_2^2 A^{\mu\nu}A_{\mu\nu}
%  -8\alpha_2^4(\cA_\mu\cA^\mu)^2
%  +2\alpha_2^2(\nabla_\mu\cA^\mu)^2 \\
%  &\quad  +\frac{4}{3}\alpha_2^2 \nabla_\mu\left(
%  \cA^\nu\nabla_\nu\cA^\mu -\cA^\mu\nabla_\nu\cA^\nu \right)\\
 &=-\frac{7}{360}R^{\mu\nu\rho\sigma}R_{\mu\nu\rho\sigma}
 -\frac{1}{45}R^{\mu\nu}R_{\mu\nu} +\frac{1}{72}R^2
 +\frac{1}{30}\dAlembert R \\
 &\quad +\frac{2}{3}\alpha_1^2 V^{\mu\nu}V_{\mu\nu}
 +\frac{2}{3}\alpha_2^2 A^{\mu\nu}A_{\mu\nu}
 -\frac{4}{3}\alpha_2^2\,\dAlembert(\cA_\mu\cA^\mu) \\
 &\quad  +\frac{4}{3}\alpha_2^2 \nabla_\mu\left(
 \cA^\nu\nabla_\nu\cA^\mu -\cA^\mu\nabla_\nu\cA^\nu \right) \\
 &\quad +\frac{1}{3}m^2R -8\alpha_2^2 m^2\cA_\mu\cA^\mu +2m^4.
\end{split}
\end{equation}
This term does not explicitly contain terms involving both curvature and
torsion. Note, however, that the term $\nabla_\mu
(\cA^\nu\nabla_\nu\cA^\mu -\cA^\mu\nabla_\nu\cA^\nu)$ contains such a
cross term between the Ricci tensor and the axial vector component of
torsion, $R_{\mu\nu}\cA^\mu\cA^\nu$.

It is interesting to compare the torsion terms of the one-loop
effective actions for a scalar field and a Dirac field. The effective
Lagrangian for a scalar field \eqref{effact0} contains the following
torsion terms:
\begin{equation}\label{tt.scalar}
\begin{split}
&P_i,\quad \dAlembert P_i,\quad (i=2,\ldots,5)\\
&P_iP_j,\quad (i=1,\ldots,5,\ j=2,\ldots5)
\end{split}
\end{equation}
where $P_i$ is defined in \eqref{P_i}. The first term in
\eqref{tt.scalar} appears both as $\Lambda^2\xi_iP_i$ and
$(\ln\Lambda/\epsilon)m^2\xi_iP_i$. The latter two terms in
\eqref{tt.scalar} appear in the term that is proportional to
$\ln(\Lambda/\epsilon)$ in the effective action \eqref{effact0}.
On the other hand, the effective Lagrangian for a Dirac
field\eqref{effact1/2} involves the following torsion terms:
\begin{equation}\label{tt.Dirac}
\begin{split}
&\cA_\mu\cA^\mu,\quad \dAlembert(\cA_\mu\cA^\mu),\quad
V^{\mu\nu}V_{\mu\nu},\quad A^{\mu\nu}A_{\mu\nu},\\
&\nabla_\mu\cA^\nu\nabla_\nu\cA^\mu,\quad (\nabla_\mu\cA^\mu)^2,\quad
R_{\mu\nu}\cA^\mu\cA^\nu.
\end{split}
\end{equation}
There are no cross terms between the three components of torsion in the
Dirac field case.
It is noteworthy that the only common torsion terms in the effective
Lagrangians for a scalar field and a Dirac field are the following
three terms involving the squared norm $\cA_\mu\cA^\mu$ of the axial
component of torsion:
\begin{equation}
\Lambda^2\cA_\mu\cA^\mu,\quad (\ln\Lambda/\epsilon)
m^2\cA_\mu\cA^\mu,\quad (\ln\Lambda/\epsilon)\dAlembert(\cA_\mu\cA^\mu).
\end{equation}
Notice that the quartic term $(\ln\Lambda/\epsilon)(\cA_\mu\cA^\mu)^2$
does not appear in the Dirac field case \eqref{trA_2.spinor}. All the
rest of the torsion terms in \eqref{tt.scalar} and \eqref{tt.Dirac}
appear only for either a scalar field or for a Dirac field.

\subsection{Induced gravitational couplings}\label{sec:couplins}
We compare the induced gravitational actions \eqref{effact0} and
\eqref{effact1/2} with the gravitational action
\begin{equation}
S=\frac{1}{16\pi G}\int d^4x\sqrt{-g}\left( R-2\lambda+\ldots \right),
\end{equation}
where the dots stand for all possible generally covariant terms
constructed from the torsion and the curvature (except $R$).
The induced cosmological constant is obtained from the induced actions
\eqref{effact0} and \eqref{effact1/2} as
\begin{equation}\label{lambda_ind}
\frac{\lambda_\mathrm{ind}}{G_\mathrm{ind}}=
\sum_f\frac{C_f^{(0)}}{4\pi}\left( \frac{\Lambda^4}{2}
-\Lambda^2m_f^2+\ln\left(\frac{\Lambda}{\epsilon}\right)m_f^4 \right),
\end{equation}
where the sum is taken over all scalar and Dirac fields ($f=s,d$), and
the constant $C_s^{(0)}=-1$ for a scalar field and $C_d^{(0)}=4$ for a
Dirac field, and the induced Newton constant $G_\mathrm{ind}$ will be
discussed below. Since the masses $m_f$ are taken to be much lighter
than the ultraviolet cutoff, $m_f\ll\Lambda$, the dominant contribution
to the cosmological constant is
\begin{equation}
\frac{\lambda_\mathrm{ind}}{G_\mathrm{ind}}\approx\frac{(4N_d-N_s)}
{8\pi} \Lambda^4,
\end{equation}
where $N_d$ and $N_s$ are the number of Dirac fields and the number
of scalar fields, respectively.
Since the ultraviolet cutoff has to be at least above the electroweak
scale, $\Lambda\gtrsim 1\,\mathrm{TeV}$, up to where the Standard Model
has been tested accurately, and setting $G_\mathrm{ind}$ to its observed
or of magnitude, $G_\mathrm{ind}\sim10^{-1}\MP^{-2}$, we obtain
$\lambda_\mathrm{ind}\lesssim-10^{-64}\MP^2$ for a scalar field. A Dirac
spinor produces a positive cosmological constant,
$\lambda_\mathrm{ind}\gtrsim10^{-64}\MP^2$.  As usual, the vacuum
energy obtained from a quantum field theory is far too high compared to
the minuscule observed value $\lambda\sim10^{-122}\MP^2$. Actually, we
will soon see that the ultraviolet cutoff is expected to be comparable
to the Planck mass in this approach, so that the prediction for
$\lambda_\mathrm{ind}$ with fermionic matter is much higher than the
above estimate, around $\lambda_\mathrm{ind}\sim\MP^2$. Sakharov's
approach does not help us with the vacuum energy problem. Consequently,
the vacuum energy term that is proportional to the volume of spacetime
is usually ignored in Sakharov's approach. It might, however, be
possible to resolve the problem with fine tuning and radiative
instability of the cosmological constant. For this purpose, we would
like to highlight the proposal of vacuum energy sequestering
\cite{Kaloper:2013zca}, and in particular its recent local formulation
\cite{Kaloper:2015jra,Bufalo:2016omb}, where the perturbative
instability of the vacuum energy contribution of quantized matter fields
is tamed by letting the gravitational and cosmological constants become
variables and introducing auxiliary volume four-forms for fixing the
values of the given constants. Large contributions to the vacuum energy
from the matter sector are cancelled by the sequestering mechanism. A
potential problem with including the vacuum energy sequestering
mechanism into Sakharov's induced gravity is that the gravitational
terms (and consequently the gravitational and cosmological constants)
are absent in the classical action prior to quantization of matter.
Hence, the additional Lagrangian that is needed for the local
sequestering mechanism has to be introduced by hand, unless one finds a
way to induce those terms along with the gravitational action. A
possible clue for such a direction might be that the gravitational
action along with the vacuum energy sequestering mechanism can be
considered to emerge from gauge fixing in an underlying theory
\cite{Bufalo:2016omb}.\footnote{In this extension of the theory
\cite{Bufalo:2016omb}, the gravitational action with the vacuum energy
sequestering mechanism is considered to be a gauge fixing action
$S_\mathrm{gf}$ for an underlying theory. The BRST-invariant action is
constructed as usual by adding the appropriate ghost action
$S_\mathrm{gh}$. The resulting gravitational action
$S_\mathrm{g}=S_\mathrm{gf}+S_\mathrm{gh}$ is found to be not only BRST
invariant but also BRST exact, in the sense that
$S_\mathrm{g}\propto\delta_BF$, where $\delta_BF$ is the BRST
transformation of a certain functional $F$.}
% For canonical formulation of the local theory of vacuum energy
% sequestering, quantization and related matters, see
% \cite{Bufalo:2016omb}.

Thus, the induced Newton constant is obtained as
\begin{equation}\label{G_ind}
\frac{1}{G_\mathrm{ind}}=\sum_f\frac{C_f^{(1)}}{\pi}\left(
\frac{\Lambda^2}{2} -\ln\left(\frac{\Lambda}{\epsilon}\right)m_f^2
\right),
\end{equation}
where $C_s^{(1)}=\frac{1}{6}-\xi_{1s}$ for a scalar field and
$C_d^{(1)}=\frac{1}{3}$ for a Dirac spinor. Since $m_f\ll\Lambda$, the
dominant contribution to the Newton constant is obtained as
\begin{equation}
\frac{1}{G_\mathrm{ind}}\approx\frac{\left( 2N_d+N_s-6\sum_s\xi_{1s}
\right)}{12\pi} \Lambda^2.
\end{equation}
This implies that for a single Dirac field $\Lambda\sim10\MP$.
Including a scalar field or scalar fields with a large negative
coupling $\xi_{1s}\ll-1$ would enable $\Lambda$ to be set lower than
$\MP$. Since such a strong nonminimal coupling of a scalar field to
the curvature is not known, and the fundamental matter fields are
fermions, we do not explore the case $\xi_{1s}\ll-1$ further here.
Thus, for a realistic model of matter, where several fermionic fields
are present, the ultraviolet cutoff $\Lambda$ is comparable to the
Planck mass.

We have obtained the one-loop effective actions for a scalar field and
for a Dirac field with arbitrary nonminimal couplings $\xi_i$ and
$\alpha_j$. In order to estimate the magnitude of the induced couplings
of the torsion terms, we have to set the magnitudes of the constants
$\xi_i$ and $\alpha_j$. For a scalar field \eqref{L_scalar} we assume
that all the couplings satisfy $|\xi_i|\lesssim1$, and the coupling to
the scalar curvature also satisfies $\xi_1\le1$, which ensures that the
induced Newton constant \eqref{G_ind} is not too negative, so that the
positive contribution from fermionic matter can outweigh it. The given
range of couplings also includes the special case
$\frac{1}{6}R-\sum_{i=1}^5\xi_i P_i=\tilde{R}$, where all couplings
$10^{-1}\lesssim|\xi_i|\lesssim1$. In the leading $\Lambda^2$ order,
that special case is Einstein--Cartain--Sciama--Kibble gravity. Since
Einstein--Cartain--Sciama--Kibble gravity resides in the upper end of
the chosen coupling range $|\xi_i|\lesssim1$, it provides an estimate
for the effect of the torsion terms in this range. The spin-spin contact
interaction in Einstein--Cartain--Sciama--Kibble gravity is weak and
becomes comparable to the effect of mass at very high mass densities
\cite{Hehl:1976kj}, around $10^{47}\;\mathrm{g/cm^3}$ for electrons and
$10^{54}\;\mathrm{g/cm^3}$ for neutrons. These densities are so high
that they are only encountered in black holes and in the early universe,
but they are still much lower than the Planck density at which the
quantum gravity effects are expected to dominate. Deviation from the
special case of Einstein--Cartain--Sciama--Kibble theory does enable
propagation of torsion, which is clearly possible in the generic
induced action, but the magnitude of the couplings remains weak.
Next we shall give a similar estimation for a Dirac field. We assume the
couplings for a Dirac field \eqref{S_1/2} satisfy $|\alpha_j|\lesssim1$.
This range includes the minimally-coupled Dirac field, which corresponds
to the couplings $\alpha_1=0$ and $\alpha_2=-\frac{1}{8}$. Thus, in the
given range of couplings, the leading-order torsion contributions in the
induced action \eqref{effact1/2} are of a similar magnitude as in the
case of the scalar field analyzed above.

In the low-energy realm of classical gravity, the curvature and torsion
terms in the contributions \eqref{trA_2.scalar} and
\eqref{trA_2.spinor}, which are multiplied by $\ln(\Lambda/\epsilon)$ in
the effective Lagrangian, are heavily suppressed compared to the leading
$\Lambda^2$ contribution discussed above. The infrared cutoff
$\epsilon$ can be chosen to be at most of the order of the mass of the
lightest matter particles, which are the neutrinos, so that we can set
$\epsilon\lesssim10^{-3}\,\mathrm{eV}$. Note that setting $\epsilon$ ten
or twenty orders of magnitude lower than $10^{-3}\,\mathrm{eV}$ would
still result in $\ln(\Lambda/\epsilon)$ being of the same order of
magnitude, so that the present discussion does not depend much on
the chosen infrared cutoff. Hence, in the coupling ranges chosen above,
the dimensionless net couplings of the higher-order terms ($Riemann^2$,
$\dAlembert R$, $Torsion^4$, $\dAlembert Torsion^2$, $(\nabla
Torsion)^2$ and $Riemann\times Torsion^2$) in the effective Lagrangian
are of the order one or below, and hence their effect on low-energy
physics is marginal (apart from their possible impact on the propagation
of torsion). This point of view can be justified by treating gravity as
a low-energy effective field theory.

\section{Induced gauge theory of gravity}\label{sec:inducedPG}
It is known that the two formulations of classical gravity on
Riemann--Cartan spacetime are not generally equivalent. In the
second-order formulation, the independent variables are the metric and
the torsion. Induced gravity in the second-order formulation was
considered in Sec.~\ref{sec:IG}, and it was noted that the induced
action can as well be written as a functional of the first-order
variables. In the gauge theory formulation, which is a first-order
formulation, the independent variables are the gauge fields required to
achieve a desired local gauge symmetry. Here we consider Poincar\'e
gauge symmetry, so that the gauge fields consists of the vierbein and
the Lorentz connection. The two formulations of gravity with both
curvature and torsion are equivalent classically for the degenerate case
of Einstein--Cartain--Sciama--Kibble gravity (see e.g
\cite{Hehl:1976kj}), but not for PG generally. The decomposition of a
Lorentz connection to a Levi-Civita connection and contortion parts can
always be inserted into the field equations of PG. Inserting the
decomposition of the connection into the action, however, changes the
theory significantly \cite{Nester:2012ib}.

Next we shall consider the induced gravitational action for PG. Recall
that gravitational dynamics does not yet exist at this stage, since the
gravitational fields play the role of (classical) background fields,
while matter fields are quantized. Only after the one-loop effective
action for matter has been obtained, we will choose the independent
variables for gravity, which will be the vierbein and the Lorentz
connection, after which the variational principle can be applied to
derive the gravitational field equations. Since the elementary object of
special relativity in the setting leading to gravity on a
Riemann--Cartan spacetime is a Dirac spinor \cite{Blagojevic:2012bc}
rather than a mass point or a scalar field, we primarily consider
quantization of Dirac fields in this section.

With the intention of obtaining an induced Poincar\'e gauge theory
of gravity, we start from the Hermitian action for a minimally-coupled
Dirac spinor,
\begin{equation}\label{S.Her.PG}
 S_{\text{Dirac}}^{\text{min.}}=\int d^4x
 \left(\det e^a_{\phantom{a}\mu}\right) \left[ \frac{i}{2}
 \left( \bar{\psi}\gamma^a\tilde\nabla_a\psi
 -\tilde\nabla_a\bar{\psi}\gamma^a\psi \right) -m\bar{\psi}\psi \right],
\end{equation}
where the Poincar\'e gauge covariant derivative is defined as
$\tilde\nabla_a\psi=e_a^{\phantom{a}\mu}\tilde\nabla_\mu\psi$ with the
definition of $\tilde\nabla_\mu\psi$ given in
\eqref{tildenabla.spinor}.\footnote{Note that in the volume element,
$\left(\det e^a_{\phantom{a}\mu}\right)=
\left(\det e_a^{\phantom{a}\mu}\right)^{-1}$, if one prefers to use
the inverse vierbein that appears in the covariant derivative.}
The action for a Dirac spinor has a similar form as in the second-order
formulation of Sec.~\ref{sec:spin1/2} with \eqref{S_1/2} or without
\eqref{S_1/2,min} nonminimal couplings, except that now the
geometry of the background is determined by the independent variables
$e^a_{\phantom{a}\mu}$ and $\tilde{\omega}_{\mu}^{\phantom{\mu}ab}$.
The minimally coupled action \eqref{S.Her.PG} is rewritten as
\begin{equation}\label{S_1/2,min.PG}
 S_{\text{Dirac}}^{\text{min.}}=\int d^4x \left(\det
 e^a_{\phantom{a}\mu}\right) \bar{\psi}\left(
 i\gamma^a\tilde\nabla_a
 -\frac{i}{2}\gamma^a\cV_a -m \right)\psi,
\end{equation}
where the vector component of torsion is defined as
$\cV_a=e_a^{\phantom{a}\mu}e_b^{\phantom{b}\nu}
T_{\nu\mu}^{\phantom{\nu\mu}b}$.
Nonminimal couplings to torsion could be included in a similar way as in
Sec.~\ref{sec:IG}. However, we should note that the principle of gauge
invariance does not require such nonminimal terms. Therefore, we will
consider the minimally coupled case for simplicity.

The one-loop effective action is defined (as in Sec.~\ref{sec:spin1/2})
as
\begin{equation}\label{effact0.PG}
 S_\mathrm{eff}^{(1)}=-\frac{i}{2}\ln\det(l^2D)
 -\frac{i}{2}\Tr\ln(l^2D),
\end{equation}
where the squared differential operator $D$ for the action
\eqref{S_1/2,min.PG} is written as
\begin{equation}\label{D.PG}
D=\left[\gamma^a\left( \tilde\nabla_a-\frac{1}{2}\cV_a \right)\right]^2
+m^2.
\end{equation}
The kernel expansion of the operator $D$ for a spinor on
Riemann-Cartain spacetime has been studied before, particularly in
\cite{Obukhov:1982da,Obukhov:1983mm}. Such calculations are
based on the decomposition of the connection into torsion-free and
torsion components in one way or another. We will adopt a similar
approach but with one crucial difference: the relation of the two
connections \eqref{omegarelation} shall be used both ways. First the
decomposition of the Lorentz connection $\tilde\omega^{ab}$ is used for
the derivation of the one-loop effective action. Next the Lorentz
connection $\tilde\omega^{ab}$ is composed back together, so that the
effective action is expressed in terms of the background fields
$e^a_{\phantom{a}\mu}$ and $\tilde\omega_\mu^{\phantom{\mu}ab}$, namely,
in terms of the gauge fields of PG. After that we can elevate the said
background fields into independent gravitational variables, and
thereafter determine their dynamics by setting up the variational
principle and deriving the field equations. Varying the action before
the full Lorentz connection is composed would lead to inequivalent field
equations \cite{Nester:2012ib}, which would ruin our chances to obtain
parity with PG. We avoid this problem with the approach described above.

The operator \eqref{D.PG} is written with \eqref{omegarelation} as
\begin{equation}\label{D.PG.2}
D=\left[\ubar{\gamma}^\mu\left( \tilde\nabla_\mu
-\frac{i}{8}\gamma_5\cA_\mu \right)\right]^2 +m^2.
\end{equation}
The one-loop effective action is derived in the same way as in
Sec.~\ref{sec:spin1/2}. Then we write it in terms of the variables
$e^a_{\phantom{a}\mu}$ and $\tilde\omega_\mu^{\phantom{\mu}ab}$.
We obtain it as
\begin{equation}\label{effact.PG}
\begin{split}
S_\mathrm{eff}^{(1)}&=-\frac{\Lambda^4}{64\pi^2}\int d^4x
\left(\det e^a_{\phantom{a}\mu}\right)\tr A_0
-\frac{\Lambda^2}{32\pi^2}\int d^4x
\left(\det e^a_{\phantom{a}\mu}\right)\tr A_1 \\
&\quad-\frac{\ln(\Lambda/\epsilon)}{16\pi^2}\int d^4x
\left(\det e^a_{\phantom{a}\mu}\right)\tr A_2
+\text{ultraviolet-finite terms}.
\end{split}
\end{equation}
The induced gravitational action at low energies is defined by the
first two terms in the effective action \eqref{effact.PG}, which are
obtained as
\begin{align}
\tr A_0&=4,\\
\tr A_1&=-\frac{1}{3}\tilde{R}
-\frac{2}{3}\tilde\nabla_a\cV^a-\frac{4}{9}\cV_a\cV^a
+\frac{1}{9}\cA_a\cA^a -\frac{1}{6}\cT_{abc}\cT^{abc} -4m^2.
\end{align}
Thus, at low energies, we obtain the induced gravitational action as
\begin{multline}\label{S_indPG}
S_\text{induced PG}^\text{low-energy}
=\frac{1}{2\kappa_\mathrm{ind}}\int d^4x\left(\det e^a_{\phantom{a}\mu}
\right) \biggl( \tilde{R} +\frac{4}{3}\cV_a\cV^a -\frac{1}{3}\cA_a\cA^a
+\frac{1}{2}\cT_{abc}\cT^{abc}\\ -2\lambda_\mathrm{ind} \biggr).
\end{multline}
This is the low-energy part of the PG action \cite{Baekler:2010fr}
with the relative coupling constants of the terms set to certain
values. Thus, we have shown that the low-energy regime of PG is induced
by quantized Dirac fields in Riemann--Cartan spacetime.\footnote{For the
nonminimally coupled Dirac action,
\[%\begin{equation}\label{S_1/2.PG}
\begin{split}
S_{\text{Dirac}}^{\text{non-min.}}&=\int d^4x \left(\det
e^a_{\phantom{a}\mu}\right)
\bar{\psi}\left( i\gamma^a\tilde\cD_a -m \right)\psi,\\
\tilde\cD_a&=\tilde\nabla_a+i\beta_1\cV_a+i\beta_2\gamma_5\cA_a,
\end{split}
\]%\end{equation}
where the coupling constants $\beta_i$ are related to the couplings
$\alpha_i$ of the action \eqref{S_1/2} as
\[%\begin{equation}
\beta_1=\alpha_1+\frac{i}{2},\quad \beta_2=\alpha_2+\frac{1}{8},
\]%\end{equation}
the induced low-energy action is the same as in Eq.~\eqref{S_indPG}
except that the contribution $-\frac{1}{3}\cA_a\cA^a$ of the axial
component of torsion is replaced by
\[
-\frac{1}{3}\left( 1-2\beta_2+8\beta_2^2 \right)\cA_a\cA^a.
\]}
Note that we have dropped the total derivative term
$\tilde\nabla_a\cV^a$ from the action \eqref{S_indPG}.
The induced gravitational couplings in the action \eqref{S_indPG} are
given as
\begin{align}
\frac{\lambda_\mathrm{ind}}{\kappa_\mathrm{ind}}
&=\frac{1}{8\pi^2}\left( \frac{N_d\Lambda^4}{2}
-\sum_d\Lambda^2m_d^2 \right),\\
\frac{1}{\kappa_\mathrm{ind}}&=\frac{N_d}{48\pi^2}\Lambda^2,
\end{align}
where $N_d$ is the number of free Dirac fields, $m_d$ is the mass of
each Dirac field, and we have omitted the terms of order
$\ln(\Lambda/\epsilon)$. What was said about the índuced couplings
in Sec.~\ref{sec:couplins} still apply, including the assessment of the
orders of magnitudes for the cutoffs.

The third term in the one-loop effective action \eqref{effact.PG}
provides the high-energy or strong-gravity regime of the induced
gravitational action. The dimensionless coupling constants of those
higher-order terms are of the order of one or lower, and hence their
effect is weak at energies well below the ultraviolet cutoff
$\Lambda\sim\MP$. The high-energy part of the induced gravitational
action \eqref{effact.PG} does not exactly match the corresponding regime
of PG. In PG, the Lagrangian is defined to be quadratic in the field
strengths, namely, in curvature and torsion. Therefore, the high-energy
part of the PG action consists of squared curvature terms. On the other
hand, the induced action typically contains all terms which are gauge
invariant and dimensionally permitted. In the present case, it means
that the high-energy part of the induced action can also include quartic
torsion terms, as well as terms which involve covariant derivatives, for
example, $\tilde\dAlembert\tilde{R}$. This is in general the case in the
effective field theory approach to gravity. One should also note that
further contributions to the high-energy part are induced at higher
loops. Thus, it is practically impossible to achieve perfect parity with
PG at high energies in the induced gravity approach.

\section{Conclusions and outlook}
We have obtained the induced gravitational action on an Einstein--Cartan
spacetime by identifying it as the cutoff-regularized one-loop
effective action of quantized matter fields. This is the generalization
of Sakharov's induced gravity \cite{Sakharov:1967pk} to a spacetime with
both curvature and torsion. When the ultraviolet cutoff $\Lambda$ is
chosen to be comparable to the Planck mass, $\Lambda\sim\MP$, the
induced Newton constant \eqref{G_ind} has the observed magnitude. As
usual, the induced cosmological constant \eqref{lambda_ind} is much
too large compared to the observed value, since the vacuum energy
contribution is comparable to the square of the ultraviolet cutoff,
$\lambda_\mathrm{ind}\sim\Lambda^2\sim\MP^2$. We speculated that it
might be possible to use the local vacuum energy sequestering mechanism
\cite{Kaloper:2015jra} in induced gravity for setting the correct value
for $\lambda_\mathrm{ind}$ and avoiding its radiative instability. In a
reasonable range of nonminimal couplings for the free matter fields, the
contribution of torsion was found to be comparable to that in
Einstein--Cartan--Sciama--Kibble gravity. Hence, the effect of torsion
is quite weak except in very high matter densities. In general, however,
the induced gravitational action is more general than the
Einstein--Cartan--Sciama--Kibble theory, which implies that propagation
of torsion is possible. In the part of the induced action that dominates
at high energies, the dimensionless coupling constants were found to be
of the order one, which implies that their effect on low-energy
physics is marginal.

Then we have set out to show that the Poincar\'e gauge theory of
gravity (PG) can be obtained by using the Sakharov induced gravity
mechanism. We have shown that the
quantization of free Dirac fields induces the low-energy part of the PG
action \eqref{S_indPG} with certain relative couplings between the
curvature and torsion terms. We conjecture that the result can be
generalized to any gauge theory of gravity, in particular to the more
general metric-affine gauge theories of gravity. The high-energy part of
the induced action was observed to differ from the Poincar\'e gauge
theory of gravity, since it does not contain only squared curvature
terms but also terms that involve covariant derivatives and further
contributions from torsion. This is to be expected in an approach based
on effective field theory, since any term that is both invariant under
the given symmetry and dimensionally allowed can, and often will,
appear in the effective action.
In conclusion, based on our derivation of the Poincar\'e gauge theory
of gravity and the general structure of the effective action, we
conjecture that the Sakharov mechanism can be used to induce the action
for any gauge theory of gravity.

If we regard that gravity is induced via quantization of matter, what
should we do with gravity itself? This is something that Sakharov's
approach cannot address. Most of us believe that gravity should be
quantized in one way or another. On the other hand, it is conceivable
that space, time and gravity could emerge at a length scale above the
Planck length, so that quantization in the conventional sense would be
unnecessary. The fundamental theory behind all that, which would not
involve a gravitational interaction, might of course be a quantum theory
of some kind. Perhaps the strongest hint towards an emergent nature of
gravity is the well known and deep connection of gravity and
thermodynamics.

The merit of Sakharov's idea in this perspective is that, having a
theory that includes a curved spacetime and also the quantized matter
and gauge fields at the sub-Planckian energies, the gravitational
interaction is necessarily produced as well. In other words, gravity
emerges as an unavoidable companion of quantum matter.

\section*{Acknowledgements}
We are much grateful to Stephen Adler, Stanley Deser, Friedrich Hehl
and Yuri Obukhov for several illuminating correspondences.
We also deeply thank Milutin Blagojevi\'c, Amir Ghalee, Yaghoub
Heydarzade, Paddy Padmanabhan, Ilya Shapiro, Andrei Smilga, Dimitri
Vassilevich and Anthony Zee for many valuable remarks.
M.O. gratefully acknowledges support from the Emil Aaltonen
Foundation.

\end{document}